\documentclass[structabstract]{aa}  
\usepackage{graphicx}
\usepackage{txfonts}
\usepackage{natbib}

\begin{document}
   \title{Different Evolutionary Stages in the Massive Star Forming Region S255 Complex}

%   \subtitle{}

   \author{Y. Wang\inst{1}\inst{,2}\inst{,3}, H. Beuther\inst{2}, A. Bik\inst{2}, T. Vasyunina\inst{2}, Z. Jiang\inst{1}, E. Puga\inst{4}\inst{,5}, H. Linz\inst{2}, J. A. Rod\'on\inst{6}, Th. Henning\inst{2}, and M. Tamura\inst{7} 
          }
   \institute{Purple Mountain Observatory, Chinese Academy of Sciences, 210008 Nanjing, China\\
              \email{ywang@pmo.ac.cn}
         \and
             Max-Plank-Institute for Astronomy, K\"onigstuhl 17, 69117
             Heidelberg, Germany\\
             \and
               Graduate University of the Chinese Academy of Sciences, 19A Yuquan Road, Shijingshan District, 100049 Beijing, China\\
           \and
           Centro de Astrobiolog{\'{i}}a (CSIC-INTA), Ctra. de Torrej\'on a Ajalvir  km-4, E-28850, Torrej\'on de Ardoz, Madrid, Spain\\
           \and
           Instituut voor Sterrenkunde, Katholieke Universiteit Leuven, Celestijnenlaan 200D, 3001 Leuven, Belgium\\
           \and
           Laboratoire dAstrophysique de Marseille, UMR6110 CNRS, 38 rue F. Joliot-Curie, 13388 Marseille, France
           \and
           	 National Astronomical Observatory of Japan, and GUAS, National Institutes of Natural Sciences, Japan\\
             }

   \date{Received 7 August 2010; accepted 11 November 2010}

% \abstract{}{}{}{}{} 
% 5 {} token are mandatory
 
  \abstract
  % context heading (optional)
  % {} leave it empty if necessary  
   {}
  % aims heading (mandatory)
   {Massive stars form in clusters, and they are often found in different evolutionary stages located close to each other. To understand evolutionary and environmental effects during the formation of high-mass stars, we observed three regions of massive star formation at different evolutionary stages that reside in the same natal molecular cloud.}
  % methods heading (mandatory)
   {The three regions S255IR, S255N and S255S were observed at 1.3 mm with the Submillimeter Array (SMA) and follow-up short spacing information was obtained with the IRAM 30m telescope. Near infrared (NIR) $H+K$-band spectra and continuum observations were taken for S255IR with VLT-SINFONI to study the different stellar populations in this region. }
  % results heading (mandatory)
   {The combination of millimeter (mm) and near infrared data allow us to characterize different stellar populations within the young forming cluster in 
   detail. While we find multiple mm continuum sources toward all regions, their outflow, disk and chemical properties vary considerably. 
   The most evolved source S255IR exhibits a collimated bipolar outflow visible in CO and H$_2$ emission, 
   the outflows from the youngest region S255S are still small and rather confined in the regions of the mm continuum peaks. 
   Also the chemistry toward S255IR is most evolved exhibiting strong emission from complex molecules, while much fewer 
   molecular lines are detected in S255N, and in S255S we detect only CO isotopologues and SO lines. 
   Also, rotational structures are found toward S255N and S255IR. Furthermore, a comparison of the NIR SINFONI and mm 
   data from S255IR clearly reveal two different (proto) stellar populations with an estimated age difference of approximately 1 Myr.}
  % conclusions heading (optional), leave it empty if necessary 
   {A multi-wavelength spectroscopy and mapping study reveals different evolutionary phases of the star formation regions. 
   We propose the triggered outside-in collapse star formation scenario for the bigger picture and the fragmentation scenario for S255IR.}

   \keywords{stars: formation -- ISM: jets and outflows -- ISM: molecules -- stars: early-type -- Hertzsprung-Russell (HR) and C-M diagrams -- stars: individual: S255IR, S255N, S255S}
 \authorrunning{Y. Wang et al.}  
\titlerunning{Different Evolutionary Stages in S255 Complex}
   \maketitle
%
%________________________________________________________________

\section{Introduction}
\label{introduction}
 Massive stars ($M>8 M_{\odot}$) are one of the paramount components in the evolution of the universe, yet their formation is significantly less well understood. than that of their low-mass counterparts. S255IR is a famous massive star formation complex at a distance of $1.59^{+0.07}_{-0.06}$ kpc \citep{rygl2010}, embraced by the Sharpless regions S255 and S257 that are already evolved H{\tiny II} regions. The SCUBA 850 $\mu$m observation \citep{difrancesco2008, klein2005} shows three main continuum sources: G192.60-MM2 \citep{minier2005} lies in the center region S255IR, and two additional mm continuum peaks G192.60-MM1 \citep{minier2005} and G192.60-MM3 \citep{minier2005} toward the northern region S255N and southern region S255S, respectively (Figure \ref{spitzer}).
   
The central region S255IR: The central IRAS source with a luminosity of $5\times10^4$ L$_{\odot}$ \citep{minier2005} harbors three UCH{\tiny II} regions region \citep{snell1986} that is associated with Class II CH$_3$OH and H$_2$O maser emission \citep{goddi2007}, which indicate the presence of massive young stellar objects. The luminosity would be  $2\times10^4$ L$_{\odot}$ if we apply the new distance of 1.59 kpc. The region hosts a cluster of low-mass sources that is surrounded by a shocked bubble of H$_2$ gas  \citep{ojha2006,miralles1997}. \citet{tamura1991} resolved the central near infrared source into two compact sources, NIRS 1 and NIRS 3. Furthermore, \citet{minier2007} reported HCO$^+$ infall signatures toward this region. In the mid-infrared, \citet{longmore2006} resolved a massive proto-binary which coincide with NIRS 1 and NIRS 3 (Figure \ref{continuum}). And NIRS 1 has been identified as a massive disk candidate by near-infrared polarization observations \citep[see also \citet{simpson2009} for a different interpretation]{jiang2008}.

The northern region S255N: This region with 10$^5$ L$_{\odot}$ \citep{minier2005} also hosts a UCH{\tiny II} region G192.584-0.041 \citep{kurtz1994} as well as Class I CH$_3$OH and H$_2$O maser emission \citep{kurtz2004, cyganowski2007}, which indicate the presence of the massive young stellar objects. The luminosity would be  $4\times10^4$ L$_{\odot}$ if we apply the new distance of 1.59 kpc. The recent Submillimeter Array observations toward that region at a spatial resolution of 3.6$^{\prime\prime}$ resolved three mm sources and strong molecular line emission, and none of the sources is associated with NIR point source \citep{cyganowski2007}. Outflow activity is evidenced by various tracers from shocked H$_2$ emission via cm continuum and SiO emission \citep{miralles1997, cyganowski2007}, and global infall was reported by \citet{minier2007}.

The southern region S255S: This sub-source is the least studied so far. It exhibits strong mm continuum emission (Figure \ref{spitzer}) but no other sign of active star formation yet such as near infrared and mid infrared emissions. Therefore, it is proposed to be in a pre-stellar phase of the evolutionary sequence \citep{minier2007}.

While G192.60-MM2 in S255IR still shows signs of active star formation, it is associated with a NIR cluster and appears to be the most evolved one of the three region. G192.60-MM1 in S255N is of similar luminosity, and G192.60-MM3 is S255S is a High-Mass Starless Core candidate. Therefore, S255 complex is the ideal candidate source to simultaneously investigate several sites of massive star formation at different evolutionary stages within the same larger-scale environment. 
%__________________________________________________________________

\section{Observations and Data reductions}

%   Two column figure (place early!)

\subsection{Submillimeter Array observations}
  The S255 complex was observed with three fields with the Submillimeter Array (SMA) on 
  November 3rd 2008 in the compact configuration with seven antennas and on February 8th, 
  February 17th 2009 in the extended configuration with eight antennas, and  February 13th in 
  the extended configuration with seven antennas. The phase centers of the fields, which are known 
  as S255IR, S255N and S255S, were R.A. 06h12m54.019s Dec. $+17^{\circ}59^{\prime}23.10^{\prime
  \prime}$ (J2000.0), R.A. 06h12m53.669s Dec. $+18^{\circ}00^{\prime}26.90^{\prime\prime}$  (J2000.0) and R.A. 06h12m56.58s Dec. $+17^{\circ}58^{\prime}32.80^{\prime\prime}$ (J2000.0), respectively. 
  The SMA has two spectral sidebands, both 2 GHz wide and separated by 10 GHz. The 
  receivers were tuned to 230.538 GHz in the upper sideband ($v_{lsr}$ = 10 km s$^{-1}$) with a 
  maximum spectral resolution of 0.53 km s$^{-1}$. The weather of February 8th and February 13th 
  was mediocre with zenith opacities $\tau$(225 GHz) larger than 0.2 measured by the Caltech Submillimeter Observatory (CSO). But the weather of November 3rd 2008 and February 17th 2009 was 
  good with zenith opacities $\tau$ (225 GHz) around 0.1. So we only used the data observed on 
  these two days. For the compact configuration on November 3rd 2008, bandpass was derived 
  from the quasar 3c454.3 observations. Phase and amplitude were calibrated with regularly 
  interleaved observations of the quasar  0530+135 (11.4$^{\circ}$ from the source). The flux calibration 
  was derived from Uranus observations, and the flux scale is estimated to be accurate within 
  20$\%$. For the extended configuration on February 17th 2009, bandpass was derived from 
  the quasar 3c273 observations. Phase and amplitude were calibrated with regularly interleaved 
  observations of the quasar 0530+135. Because of the lack of flux calibrator observations, the flux was 
  estimated by the SMA calibrator database for the gain calibrator, to be accurate within 20$\%$. We 
  merged the two configuration data set together, applied different robust parameters for the 
  continuum and line data, and got the synthesized beam sizes between 1.4$^{\prime\prime}$ $\times$1.1$^{\prime\prime}$ (PA 86$^\circ$) and 1.9$^{\prime\prime}\times1.6^{\prime\prime}$ (PA 
  77$^\circ$), respectively. The 3$\sigma$ rms of 1.3 mm continuum image is $\sim$ 4.5 mJy and the 3$\sigma$ rms 
  of the line data is 0.14 Jy/beam at 2 km s$^{-1}$ spectral resolution. The flagging and 
  calibration was done with the IDL superset MIR \citep{scoville1993} which was originally 
  developed for the Owens Vally Radio Observatory and adapted for the SMA\footnote{The MIR 
  cookbook by Chunhua Qi can be found at http://cfa-www.harvard.edu/$\sim$cqi/mircook.html.}. 
  The imaging and data analysis were conducted in MIRIAD \citep{sault1995}.

\subsection{Short spacing from the IRAM 30 m}
 To complement the CO$(2-1)$ observations with the missing short spacing information and to 
 investigate the large-scale general outflow properties to find the connection of the three regions, 
 we observed them with the HERA array at the IRAM 30 m telescope on November 10th 2009. 
 The $^{12}$CO$(2-1)$ line at 230.5 GHz and $^{13}$CO$(2-1)$ line at 220.4 GHz were observed in the
 on-the-fly mode. We mapped the whole region with a size of 5$^{\prime} \times$ 3$^{\prime}$ 
 centered at R.A. 06h12m54.02s Dec. $+17^{\circ}59^{\prime}23.10^{\prime\prime}$ (J2000.0). The 
 sampling interval was 3.5$^{\prime\prime}$, a bit better than Nyquist-sampling to minimize beam-smearing effects. 
 The region was scanned two times in the declination and right 
 ascension direction, respectively, in order to reduce effects caused by the scanning process. The 
 spectra were calibrated with CLASS which is part of the GILDAS software package, then the 
 declination and right ascension scans were combined with the plait algorithm in GREG which is 
 another component of the GILDAS package. The $^{12}$CO data has a beam size of 11$^{\prime\prime}$, and the rms noise level of the corrected $T_{mb}$ is around 0.23 K at 0.6 km s$^{-1}$ 
 spectral resolution, the $^{13}$CO has the same beam size but the rms noise is 0.12 K at 0.6 km s$^{-1}$ spectral resolution.

 After reducing the 30m data separately, single dish $^{12}$CO and $^{13}$CO data were 
 converted to visibilities and then combined with the SMA data using MIRIAD package task 
 UVMODEL. With different uv-range selections, the synthesized beam of the combined data varies 
 from $2.1^{\prime\prime}\times1.8^{\prime\prime}$ (PA $-84^{\circ}$) to $11.1^{\prime\prime}\times10.7^{\prime\prime}$ (PA $-74^\circ$).

\subsection{VLT-SINFONI integral field spectroscopy observations}
Near Infrared $H$- and $K$-band observations were performed using the Integral Field instrument 
SINFONI \citep{eisenhauer2003, bonnet2004} on UT4 (Yepun) of the VLT at Paranal, 
Chile. The observations of S255IR were performed in service mode on February 8th, February 12th, February 13th and March 29th 2007. The non-AO mode of SINFONI, used in combination with the setting providing the widest field of view ($8^{\prime\prime}\times8^{\prime\prime}$) with a spatial resolution of 0.25$^{\prime\prime}$ per slitlet. The typical seeing during the observations was 0.7$^{\prime\prime}$ in $K$-band. The H+K grating was used, covering these bands with a resolution of R=1500 in a single exposure.

S255IR was observed with a detector integration time (DIT) of 30 seconds per pointing. The observations were centered on the central near infrared source NIRS 1 at coordinates: R.A. 06h12m53.85s Dec. $+17^{\circ}59^{\prime}23.71^{\prime\prime}$ (J2000.0). We observed this area with SINFONI using a raster pattern, covering every location in the cluster at least twice, resulting in an effective integration time of 60 seconds per location in the field. The offset in the east-west direction was 4$^{\prime\prime}$ (i.e. half the FOV of the detector) and the offset in the north-south direction was 6.75$^{\prime\prime}$, results a total observation field of 70$^{\prime\prime}\times70^{\prime\prime}$. A sky frame was taken every 3 minutes using the same DIT as the science observations. The sky positions were chosen based on existing NIR imaging in order to avoid contamination. Immediately after every science observation, a telluric standard star was observed, matching as close as possible the airmass of the object.

The data were reduced using the SPRED (version 1.37) software developed by the MPE SINFONI 
consortium \citep{schreiber2004, abuter2006}. The procedure described by 
\citet{davies2007} was applied to remove the OH line residuals. To calibrate the flux and 
remove the telluric absorption lines, the extracted spectrum of one standard star in each OB (OB1: Hip064656, OB2: Hip048128, OB3: Hip032108, OB4: Hip031899) is used. The flux calibration of the spectra uses the 2MASS \citep{cohen2003} magnitude of the standard stars (see \citet{bik2010} for the details of the data reduction).

\section{Results}
\label{results}
\subsection{Millimeter continuum emission}
\begin{figure}[htbp]
   \centering
    \includegraphics[angle=-90,width=8cm]{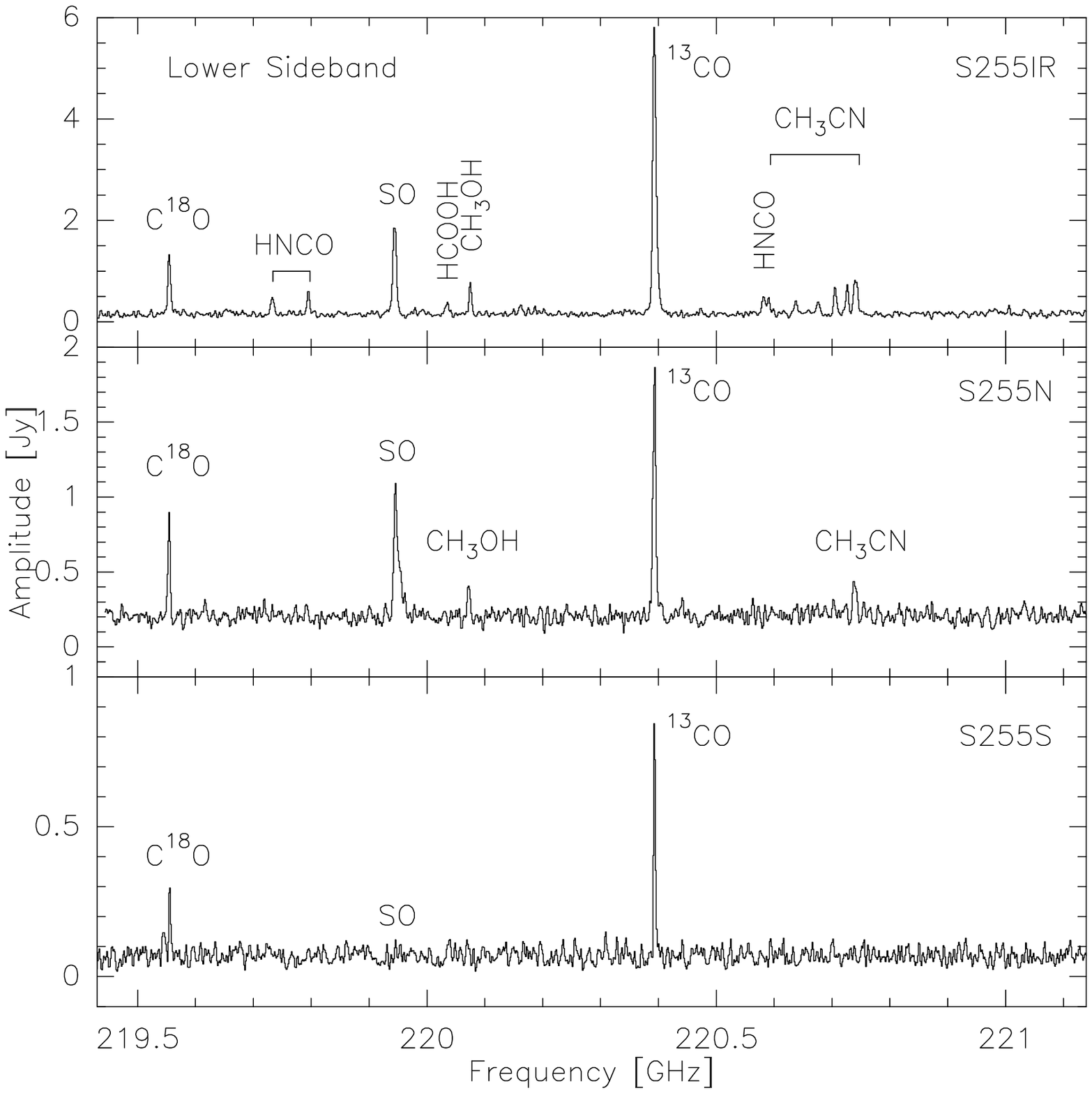}
    \includegraphics[angle=-90,width=8cm]{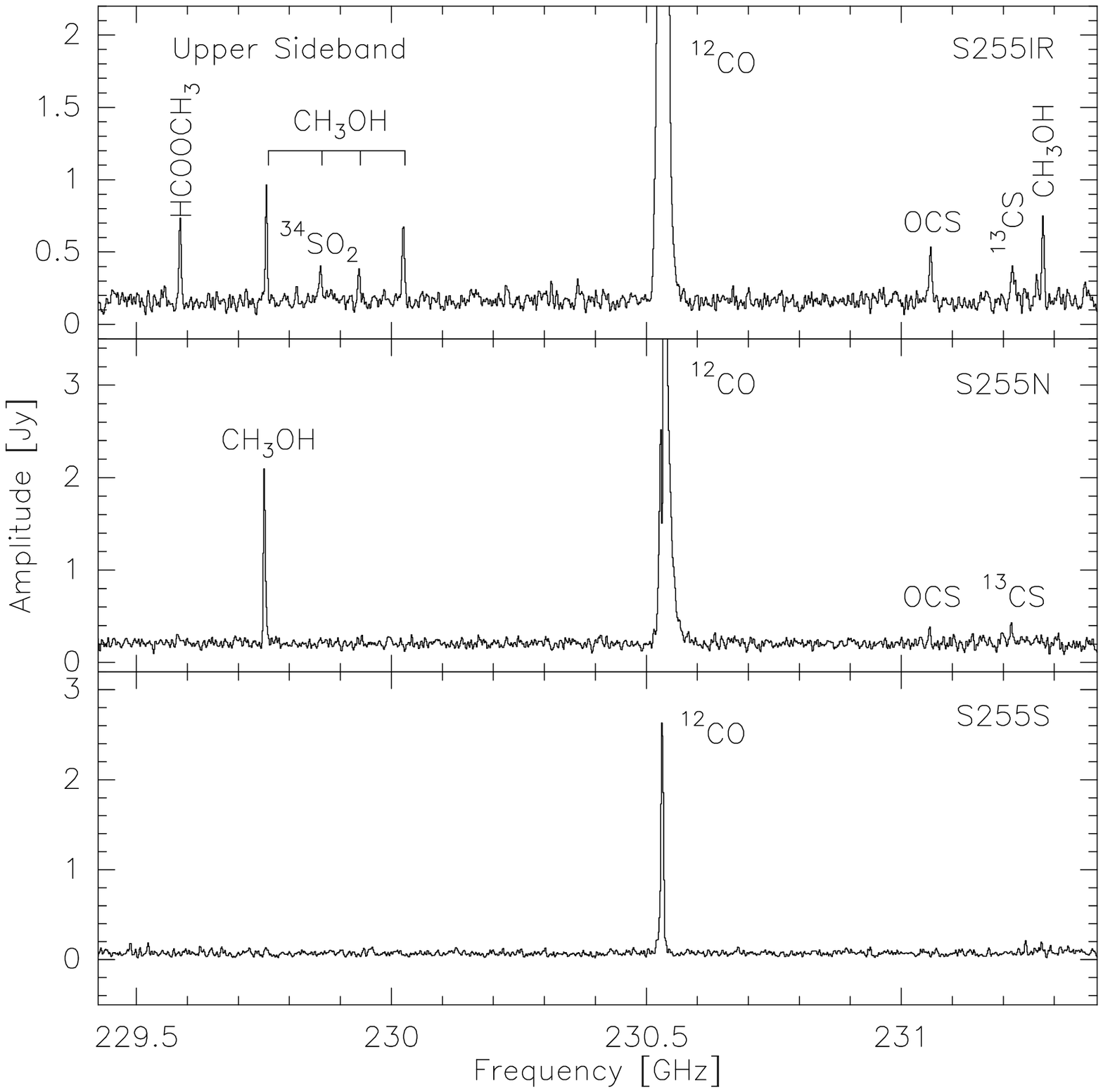}
    \caption{Lower and Upper sideband spectral vector-averaged over all baselines with a resolution 
    of 2 km s$^{-1}$ per channel. In the Upper sideband image, the $^{12}$CO line in S255IR and 
    S255N panels was not fully plotted, which goes to 11.6 Jy and 5.9 Jy respectively.}
\label{spectra}
\end{figure}

\begin{figure*}[htbp]
   \begin{center}
    \includegraphics[angle=-90,width=17.6cm]{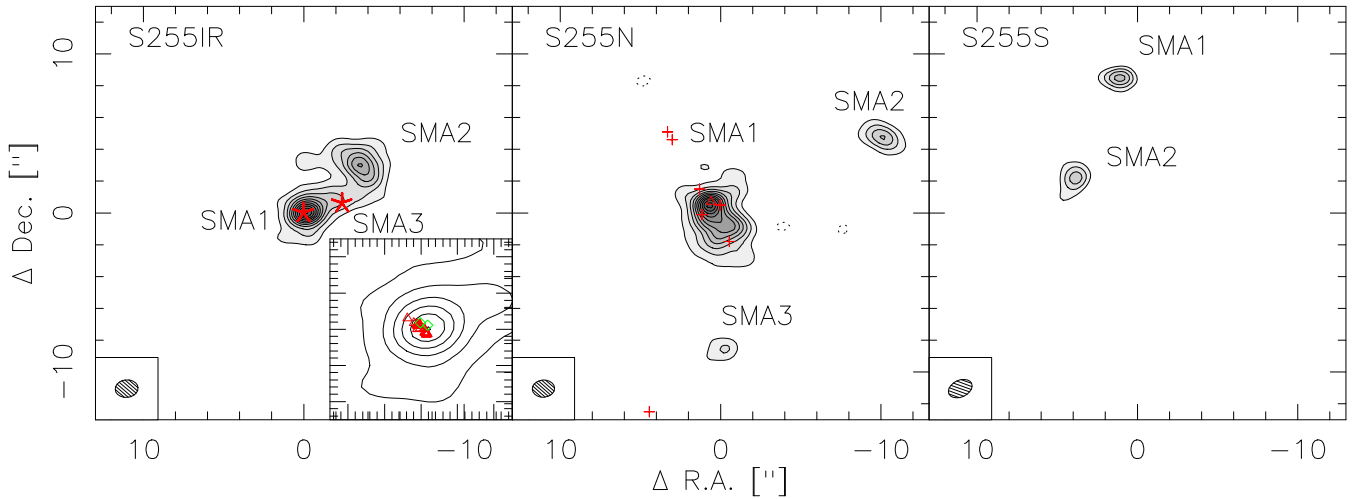}
    \caption{The SMA 1.3 mm continuum map of S255IR, S255N and S255S.
      contour levels start at 5$\sigma$ with a step of 5$\sigma$ in S255IR (1$\sigma=1.7$ mJy/beam)
      and S255N (1$\sigma=1.6$ mJy/beam) image and 3$\sigma$ in S255S (1$\sigma=0.9$ mJy/beam) image.
     The dotted contours are the negative features due to the missing flux. The synthesis beam is shown at the bottom left of each figure.
    In the S255IR panel, the asterisk marks the position of NIRS 3 and the right one marks NIRS 1 \citep{tamura1991}. The inset is a zoom in the continuum image of the inner region of S255IR-SMA1 and the contour levels start at 5$\sigma$ with a step of 10$\sigma$. The triangles are the water masers \citep{goddi2007} and the two open diamond mark the positions of two 6.7 GHz methanol masers detected by \citet{xu2009}. In S255N panel, the crosses mark the position of the Class I 44 GHz ($7_0-6_1$) A$^+$ methanol masers detected by \citet{kurtz2004} and the triangle is the water maser \citep{cyganowski2007}. The (0,0) point in each panel from left to right is R.A. 06h12m54.019s Dec. $17^{\circ}59^{\prime}23.10^{\prime\prime}$ (J2000.0), R.A. 06h12m53.669s Dec. $+18^{\circ}00^{\prime}26.90^{\prime\prime}$  (J2000.0) and R.A. 06h12m56.58s Dec. $+17^{\circ}58^{\prime}32.80^{\prime\prime}$ (J2000.0), respectively.}
\label{continuum}
\end{center}
\end{figure*}

We averaged the apparently line-free parts of upper and  lower sideband spectra of each region 
(presented in Figure \ref{spectra}) and got the continuum images of the three regions shown in Figure \ref{continuum}.

Assuming optical thin dust emission, we estimated the gas mass and column density of the continuum sources following the equations outlined in \citet{hildebrand1983} and \citet{beuther2005b}. We assumed a dust temperature of 40 K, grain size of 0.1 $\mu$m, grain mass density of 3 g cm$^{3}$, gas-to-dust ratio of 100 and a grain emissivity index of 2 (corresponding to $\kappa\approx0.3$ for comparison of \citet{ossenkopf1994}). The results are shown in Table \ref{conttable}. With the same set-up, we also calculated the gas mass and the column density of the SCUBA 850 $\mu$m \citep{difrancesco2008, klein2005} continuum peaks shown in Table \ref{scubacont}. To get the impression of how much flux we lost in the interferometer observations, we produced the SMA continuum map of the each region with the same beam size as the SCUBA 850 $\mu$m map (14$^{\prime\prime}\times14^{\prime\prime}$, \citet{difrancesco2008}). We measured the continuum flux, calculated the gas mass and compare the results with the ones in Table \ref{scubacont}; the mass we obtained from the SMA observations for S255IR, S255N and S255S is only 7.8$\%$, 8.7$\%$ and 2$\%$ of the SCUBA 850 $\mu$m measurements, respectively. Since our interferometer observations are not sensitive to spatial scales $>$ 24$^{\prime\prime}$ (the shortest base line =8.5 k$\lambda$), we filter out the smoothly distributed large-scale halo and left only the compact cores. Also, the fact that we filtered out more flux for the youngest region S255S indicates that at the earliest evolutionary stages the gas is more smoothly distributed than during later stages where the collapse produces more centrally condensed structures. Nevertheless, all derived column densities are of the order or above the proposed threshold for high-mass star formation of 1 g cm$^{-2}$ \citep{krumholz2008}.

\begin{figure}[htbp]
   \centering
    \includegraphics[angle=-90,width= 8cm]{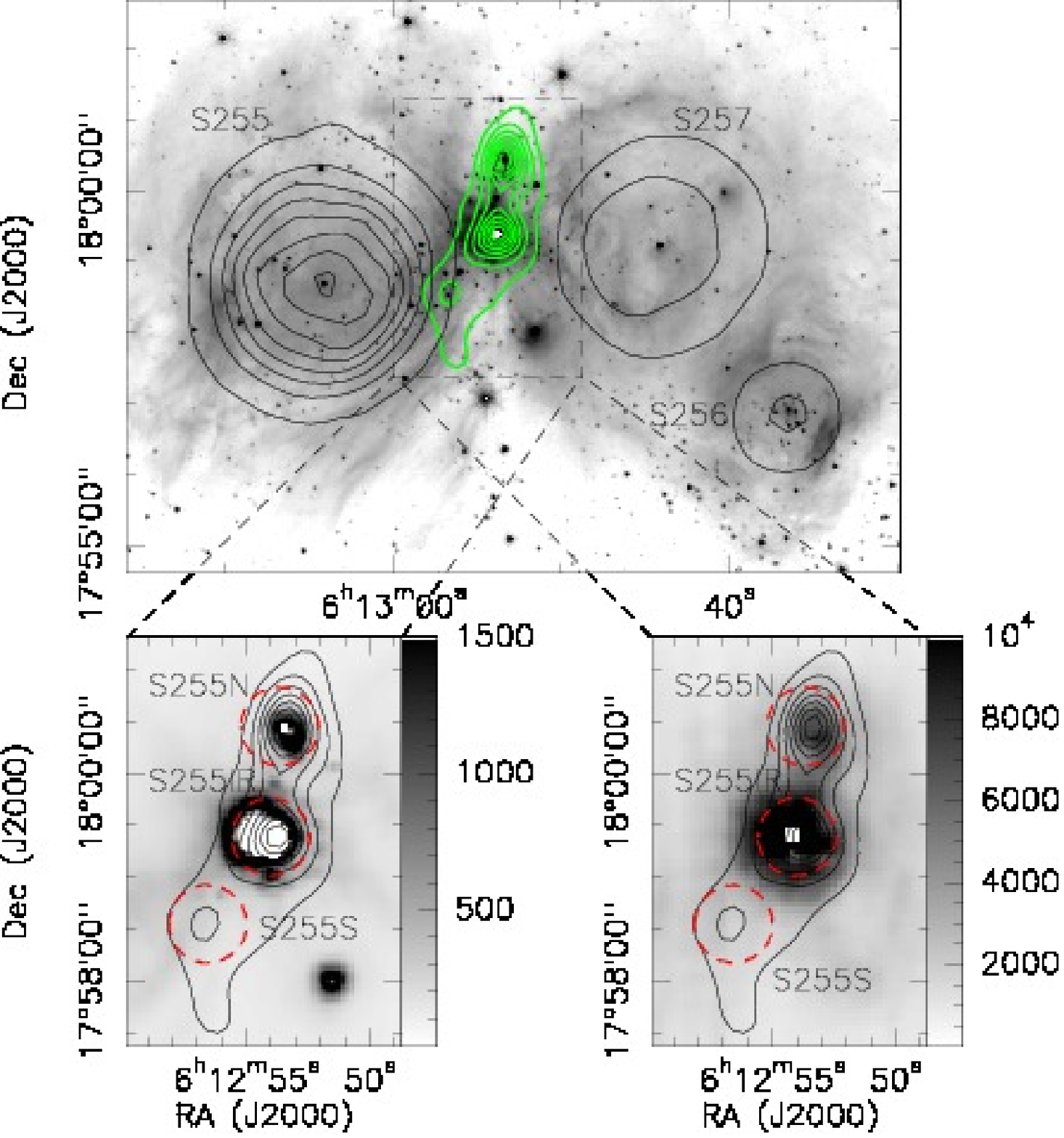}
    \caption{SPITZER IRAC 4.5 ${\mu}$m image \citep{chavarria2008} overlaid with SCUBA 850 $\mu$m contours \citep{difrancesco2008, klein2005}.  
    In the upper panel, the grey scale is 
    the SPITZER IRAC 4.5 ${\mu}$m image, the black contours are the NVSS 1.4 GHz emission and the 
    green contours are the SCUBA 850 $\mu$m continuum. In the bottom left panel, the grey scale is 
    the SPITZER MIPS 24 ${\mu}$m image, the contours are the SCUBA 850 ${\mu}$m continuum, the 
    three dashed circles mark the primary beam of our SMA observations. In the bottom right panel, 
    the grey scale is the SPITZER MIPS 70 $\mu$m image, the contours and the circles are the same as 
    the ones in the left  panel. The contour levels of the NVSS start at 10$\sigma$ (1$\sigma=0.6$ mJy beam$^{-1}$) with a step of 
    5$\sigma$. For the SCUBA 850 $\mu$m, the contour levels start at 10$\sigma$ (1$\sigma=0.1$ Jy beam$^{-1}$) with a 
    step of 10$\sigma$. The SPITZER/IRAC post-bcd data processed with pipeline version S18.7.0 and the MIPS post-bcd data processed with pipeline version S17.2.0 have been downloaded from the SPITZER archive to create these images.}
\label{spitzer}
\end{figure}

\begin{table*}
\caption{Millimeter continuum peaks properties.}
\label{conttable}
\centering
\begin{tabular}{lccrrrr}
\hline \hline
\noalign{\smallskip}
Source   & R.A. & Dec. &$S_{peak}$ & S$_{int.}$ & Mass &N$_{H_2}$\\
       & (J2000.0)& (J2000.0) &mJy/beam & mJy& [$M_\odot$] & cm$^{-2}$  \\
\hline
\noalign{\smallskip}
   S255IR-SMA1 & 06:12:54.01&+17:59:23.1 & 95 &171 & 12 & 5.5 $\times$ 10$^{24}$\\
   S255IR-SMA2 & 06:12:53.77& +17:59:26.1& 52 &150 & 10 & 3.0 $\times$ 10$^{24}$\\
   S255IR-SMA3 & 06:12:53.88& +17:59:23.7& 30 & 30  & 2  & 1.7 $\times$ 10$^{24}$\\
   S255N-SMA1  & 06:12:53.71& +18:00:27.3& 91 &327 & 22 & 5.5 $\times$ 10$^{24}$\\
   S255N-SMA2  & 06:12:52.95& +18:00:31.7& 33 &47 & 3 & 2.0 $\times$ 10$^{24}$\\
   S255N-SMA3  & 06:12:53.65& +18:00:18.3& 18 &19 & 1 & 1.1 $\times$ 10$^{24}$\\
   S255S-SMA1  & 06:12:56.65& +17:58:41.2& 14 &17 & 1 & 7.7 $\times$ 10$^{23}$\\
   S255S-SMA2  & 06:12:56.85& +17:58:35.0& 12 &15 & 1 & 6.6 $\times$ 10$^{23}$\\
\hline
\end{tabular}
\end{table*}

\begin{table}
\caption{SCUBA sub-millimeter continuum data.}
\label{scubacont}
\centering
\begin{tabular}{lrrrr}
\hline \hline
\noalign{\smallskip}
Source   & S$_{peak}$ & S$_{int}$ &  Mass &N$_{H_2}$\\
       & Jy/beam & Jy &[$M_\odot$] & cm$^{-2}$  \\
\hline
\noalign{\smallskip}
   S255IR &7.81 & 31 & 372 & 6.5 $\times$ 10$^{23}$\\
   S255N &7.75 & 30 & 357 & 6.4 $\times$ 10$^{23}$\\
   S255S  &2.33 & 9 & 110 & 1.9 $\times$ 10$^{23}$\\
\hline
\end{tabular}
\end{table}

\paragraph{S255IR}
The continuum image of S255IR is presented in the left panel of Figure \ref{continuum}. We resolved two main continuum peaks (i.e. S255IR-SMA1 and S255IR-SMA2) and one unresolved sub peak (i.e. S255IR-SMA3) in this region. The stronger peak named as S255IR-SMA1 in the southeast coincides with the near infrared source NIRS 3 \citep{tamura1991}. S255IR-SMA1 coincides with a UCH{\tiny II} region \citep{snell1986} generated by NIRS 3, and the inset in the S255IR panel in Figure \ref{continuum} shows that it is also associated with Class II CH$_3$OH and H$_2$O maser emissions \citep{goddi2007, minier2000, minier2005}. An unresolved peak S255IR-SMA3 that coincides with the near infrared source NIRS 1 \citep{tamura1991}, which lies 2.4$^{\prime\prime}$ west to S255IR-SMA1 and has been identified as a massive disk candidate by NIR polarization observation \citep{jiang2008}. These two continuum sources are both respectively coincident with the mid-infrared massive proto-binary source 1 and 2 identified by \citet{longmore2006}. S255IR-SMA2, which has never been detected before, is not associated with any infrared source and likely to be in a very young source, it also shows only a few lines, which will be discussed in detail in Sec. \ref{Spectral}. 

\paragraph{S255N}
The middle panel of Figure \ref{continuum} is the continuum image of S255N. The strongest peak named as S255N-SMA1 which coincides with the UCH{\tiny II} region G192.584-0.041 \citep{kurtz1994} is associate with Class I 44 GHz ($7_0-6_1$) A$^+$ methanol masers which are detected by \citet{kurtz2004} and water maser \citep{cyganowski2007}. All the methanol masers which are marked with crosses in the S255N panel in Figure \ref{continuum} are distributed along the direction of the outflow (Figure \ref{s255noutcom}, see Sec. \ref{outflow}). S255N-SMA1 is also associated with centimeter continuum emission \citep{cyganowski2007} which shows an elongation aligned with the outflow (Figure \ref{s255noutcom}, left panel). However, all three continuum sources in S255N have no corresponding near infrared point sources.

\paragraph{S255S}
The right panel of Figure \ref{continuum} presents the continuum image of S255S. There are no near 
infrared sources or mid infrared peaks (SPITZER MIPS 24 $\mu$m and 70 $\mu$m, Figure \ref{spitzer}) that are associated with these two sources. Meanwhile, the gas mass is only 2$\%$ of the SCUBA 850 $\mu$m measurements, which indicates that the gas in this region is at an early stage and smoothly distributed. All these features indicate that S255S is an extremely young region.

\subsection{Spectral line emission}
\label{Spectral}
The observed spectra of lower sideband (LSB) and upper sideband (USB) of three region are shown in Figure \ref{spectra}. 

\begin{table}
\caption{Observed lines in S255IR.}
\label{s255irlines}
\centering
\begin{tabular}{lrr}
\hline \hline
\noalign{\smallskip}
$\nu$      & line     &  {\it E}$_{\rm lower}/h$   \\
$[$GHz$]$ &~         & $[$K$]$ \\
\hline
\noalign{\smallskip}
   LSB     &~   &~ \\
\hline
\noalign{\smallskip}
   219.560  &C$^{18}$O(2$-$1)             &      5.3 \\
   219.734  &HNCO(10$_{2,9}$$-$9$_{2,8}$)  &      219  \\
   219.798  &HNCO(10$_{0,10}$$-$9$_{0,9}$)    &      48 \\
   219.949  &SO(6$_{5}$$-$5$_{4}$)           &     24 \\
   220.038  &HCOOH(10$_{0,10}$$-$9$_{0,9}$)   &    28.2\\
   220.079  &CH$_3$OH(8$_{0,8}$$-$7$_{1,6}$)E  &     85 \\
   220.339  &$^{13}$CO(2$-$1)                &     5.3 \\
   220.585  &HNCO(10$_{1,9}$$-$9$_{1,8}$)     &     91 \\  
   220.594  &CH$_3$CN(12$_6$$-$11$_6$)       &      315 \\
   220.641  &CH$_3$CN(12$_5$$-$11$_5$)       &      237 \\
   220.679  &CH$_3$CN(12$_4$$-$11$_4$)       &      173 \\
   220.709  &CH$_3$CN(12$_3$$-$11$_3$)       &      123 \\
   220.730  &CH$_3$CN(12$_2$$-$11$_2$)       &      87 \\
   220.743  &CH$_3$CN(12$_1$$-$11$_1$)       &      65 \\
   220.747  &CH$_3$CN(12$_0$$-$11$_0$)       &      58 \\
\hline
\noalign{\smallskip}
   USB       &~   &~\\
\hline
\noalign{\smallskip}
   229.590  &HCOOCH$_3$(19$_{3,16}$$-$18$_{4,15}$)E    &   106 \\
   229.759  &CH$_3$OH(8$_{-1,8}$$-$7$_{0,7}$)E          &   77 \\
   229.858  &$^{34}$SO$_2$(4$_{2,2}$$-$3$_{1,3}$)       &   7.7 \\
   229.864  &CH$_3$OH(19$_{5,15}$$-$20$_{4,16}$)A$+$    &   568 \\
   229.939  &CH$_3$OH(19$_{5,14}$$-$20$_{4,17}$)A$-$    &   568 \\
   230.027  &CH$_3$OH(3$_{2,2}$$-$4$_{1,4}$)E           &   28  \\
   230.538  &$^{12}$CO(2$-$1)                          &   5.5  \\
   231.061  &OCS(19$-$18)                              &   100  \\
   231.221  &$^{13}$CS(5$_0$$-$4$_0$)                   &   22  \\
   231.281  &CH$_3$OH(10$_{2,9}$$-$9$_{3,6}$)A$-$       &  98.8 \\
\hline

\end{tabular}
\end{table}

\begin{table}
\caption{Observed lines in S255N.}
\label{s255nlines}
\centering
\renewcommand{\footnoterule}{}  % to avoid a line before footnotes
\begin{tabular}{lrr}
\hline \hline
\noalign{\smallskip}
$\nu$      & line     &  {\it E}$_{\rm lower}/h$   \\
$[$GHz$]$ &~         & $[$K$]$ \\
\hline
\noalign{\smallskip}
   LSB     &~   &~ \\
\hline
\noalign{\smallskip}
   219.560  &C$^{18}$O(2$-$1)             &      5.3 \\
   219.949  &SO(6$_{5}$$-$5$_{4}$)           &     24 \\
   220.079  &CH$_3$OH(8$_{0,8}$$-$7$_{1,6}$)E  &     85 \\
   220.339  &$^{13}$CO(2$-$1)                &     5.3 \\
   220.743  &CH$_3$CN(12$_1$$-$11$_1$)       &      65 \\
   220.747  &CH$_3$CN(12$_0$$-$11$_0$)       &      58 \\
\hline
\noalign{\smallskip}
   USB       &~   &~\\
\hline
\noalign{\smallskip}
   229.759  &CH$_3$OH(8$_{-1,8}$$-$7$_{0,7}$)E          &   77 \\
   230.538  &$^{12}$CO(2$-$1)                          &   5.5  \\
   231.061  &OCS(19$-$18)                              &   100  \\
   231.221  &$^{13}$CS(5$_0$$-$4$_0$)                   &   22  \\
\hline

\end{tabular}
\end{table}
\begin{table}
\caption{Observed lines in S255S.}
\label{s255slines}
\centering
\renewcommand{\footnoterule}{}  % to avoid a line before footnotes
\begin{tabular}{lrr}
\hline \hline
\noalign{\smallskip}
$\nu$      & line     &  {\it E}$_{\rm lower}/h$   \\
$[$GHz$]$ &~         & $[$K$]$ \\
\hline
\noalign{\smallskip}
   LSB     &~   &~ \\
\hline
\noalign{\smallskip}
   219.560  &C$^{18}$O(2$-$1)             &      5.3 \\
   219.949  &SO(6$_{5}$$-$5$_{4}$)           &     24 \\
   220.339  &$^{13}$CO(2$-$1)                &     5.3 \\
\hline
\noalign{\smallskip}
   USB       &~   &~\\
\hline
\noalign{\smallskip}
   230.538  &$^{12}$CO(2$-$1)                          &   5.5  \\
\hline
\end{tabular}
\end{table}

\paragraph{S255IR}
In S255IR, we detect 25 lines from 10 species. Besides three normal CO isotopologues, we also detect some sulfur-bearing species (SO, $^{34}$SO$_2$, OCS, $^{13}$CS) and some dense gas molecules which are usually used to trace high mass hot cores, such as CH$_3$OH, CH$_3$CN and HCOOCH$_3$ \citep{nomura2004, beuther2009, sutton1985}. All the lines we detect are with lower energy levels $E_{lower}/k$ between 5.3 to 568 K (Table \ref{s255irlines}).

\begin{figure*}[htbp]
   \centering
    \includegraphics[angle=-90,width=\textwidth]{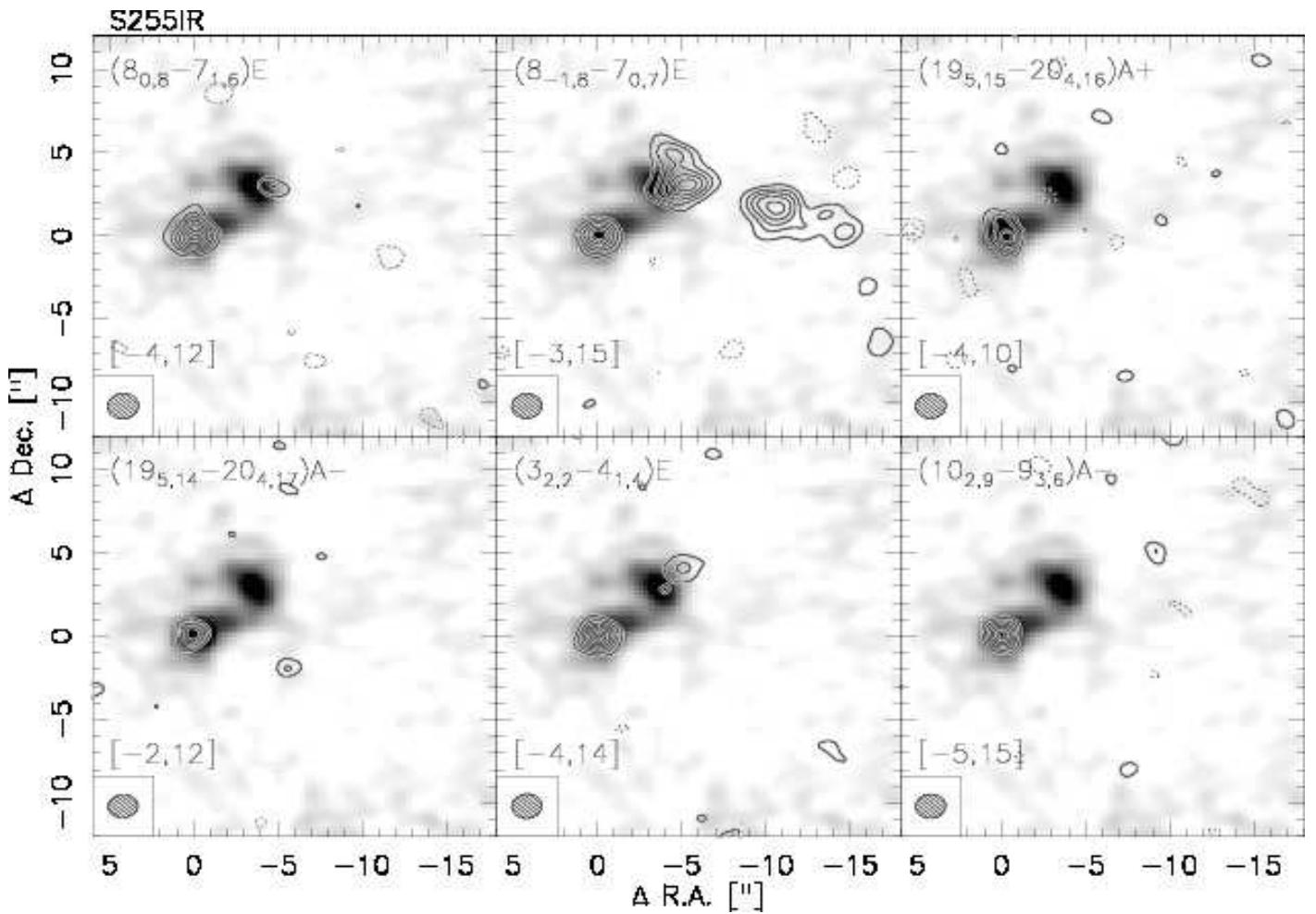}
    \caption{S255IR CH$_3$OH line integrated intensity images with the SMA 1.3 mm continuum emission in the background. 
      Contour levels start at 3$\sigma$ with 2$\sigma/$level. 
   For the top panels, the $\sigma$ of the contours in each panel, from left to right, is 12, 15, 11 mJy beam$^{-1}$, respectively, and for the bottom panels is 12, 14, 15 mJy beam$^{-1}$, respectively. 
   The dotted contours are the negative feature. 
   The integral velocity ranges are shown at the bottom-left of each panel in km s$^{-1}$.
   The synthesized beams are shown at the bottom-left of each panel. 
   The (0,0) point in each panel is R.A. 06h12m54.019s Dec. $+17^{\circ}59^{\prime}23.10^{\prime\prime}$ (J2000.0).}
\label{irch3oh}
\end{figure*}

\begin{figure*}[htbp]
   \centering
    \includegraphics[angle=-90,width=\textwidth]{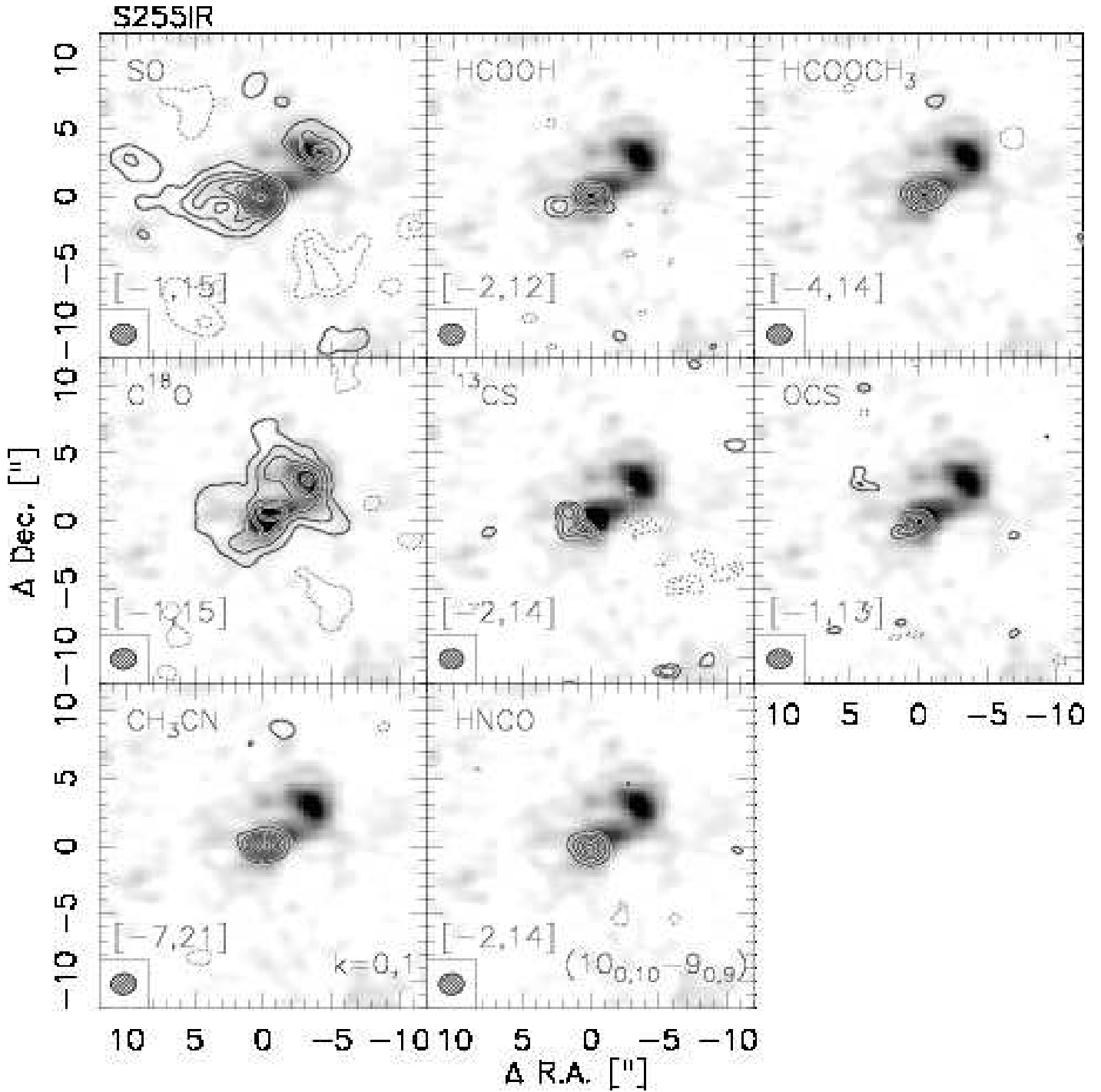}
    \caption{S255IR molecular line integrated intensity contour
      with the SMA 1.3 mm continuum emission in the background. 
     All the contour levels start at 3 $\sigma$. The contour step is
     3 $\sigma$ in SO (1$\sigma=22$ mJy beam$^{-1}$), HCOOH (1$\sigma=10$ mJy beam$^{-1}$), HCOOCH$_3$ (1$\sigma=12$ mJy beam$^{-1}$), C$^{18}$O (1$\sigma=16$ mJy beam$^{-1}$), CH$_3$CN (1$\sigma=11$ mJy beam$^{-1}$) and HNCO (1$\sigma=13$ mJy beam$^{-1}$) images, 
     and 1 $\sigma$ in OCS (1$\sigma=13$ mJy beam$^{-1}$)and $^{13}$CS(1$\sigma=11$ mJy beam$^{-1}$) image.
     The dotted contours are the negative features with the same contours as the positive ones ini each panel. 
      The integral velocity ranges are shown at the bottom-left of each panel in km s$^{-1}$.
      The synthesized beams are shown at the bottom left of each panel. We only plotted the $k=0\&1$ line for CH$_3$CN and $(10_{0,10}-9_{0,9})$ line for HNCO. 
      The (0,0) point in each panel is R.A. 06h12m54.019s Dec. $+17^{\circ}59^{\prime}23.10^{\prime\prime}$ (J2000.0).}
\label{irothers}
\end{figure*}

Figures \ref{irch3oh} and \ref{irothers} present the integrated line images of all species (except $^{34}$SO$_2$ which is blended with CH$_3$OH and too weak for imaging, and $^{12}$CO and $^{13}$CO will be discussed in details in \ref{outflow}) with different transitions. Most lines show compact emission peaked at S255IR-SMA1, and only a few show emission at S255IR-SMA2. Several CH$_3$OH lines (CH$_3$OH(8$_{0,8}-7_{1,6}$)E, CH$_3$OH(8$_{-1,8}-7_{0,7}$)E and CH$_3$OH(3$_{2,2}-4_{1,4}$)E, in Figure \ref{irch3oh}) exhibit extended emission towards S255IR-SMA2. The CH$_3$OH(8$_{-1,8}-7_{0,7}$)E line even extends to the northwest of S255IR-SMA1 following the direction of the blue-shifted outflow of S255IR-SMA1 (Figure \ref{s255iroutcom}, see Sec. \ref{outflow}). We suggest that particular the (8$_{-1,8}-7_{0,7}$)E line is due to the shock heating excited by the outflow, which could also be mixed up with some Class I methanol maser emission \citep{sutton2004, sobolev2005, kalenskii2002, slysh2002}. SO line emission associated with S255IR-SMA1 is elongated along the outflow and could be affected by the outflow. $^{13}$CS line emission forms a shell around S255IR-SMA1 (Figure \ref{irothers}, similar shell-like C$^{34}$S emission has also been found by \citet{beuther2009}). Only a few lines are associated with S255IR-SMA2, SO, C$^{18}$O (Figure \ref{irothers}) and several methanol lines (Figure \ref{irch3oh}). S255IR-SMA2 appears to be chemically younger than S2555IRmm1. We extracted C$^{18}$O spectra from these two continuum peaks. The spectrum toward S255IR-SMA1 shows two peaks and can not be fitted by a single Gaussian profile,  therefore, we calculated the FWZI (full width at zero intensity) of these two C$^{18}$O spectra, which is 12 km s$^{-1}$ and 6.8 km s$^{-1}$ for S255IR-SMA1 and S255IR-SMA2, respectively. The difference of the line width also suggests that S255IR-SMA2 is at a younger evolutionary stage.

\begin{figure*}[htbp]
   \centering
    \includegraphics[angle=-90,width=\textwidth]{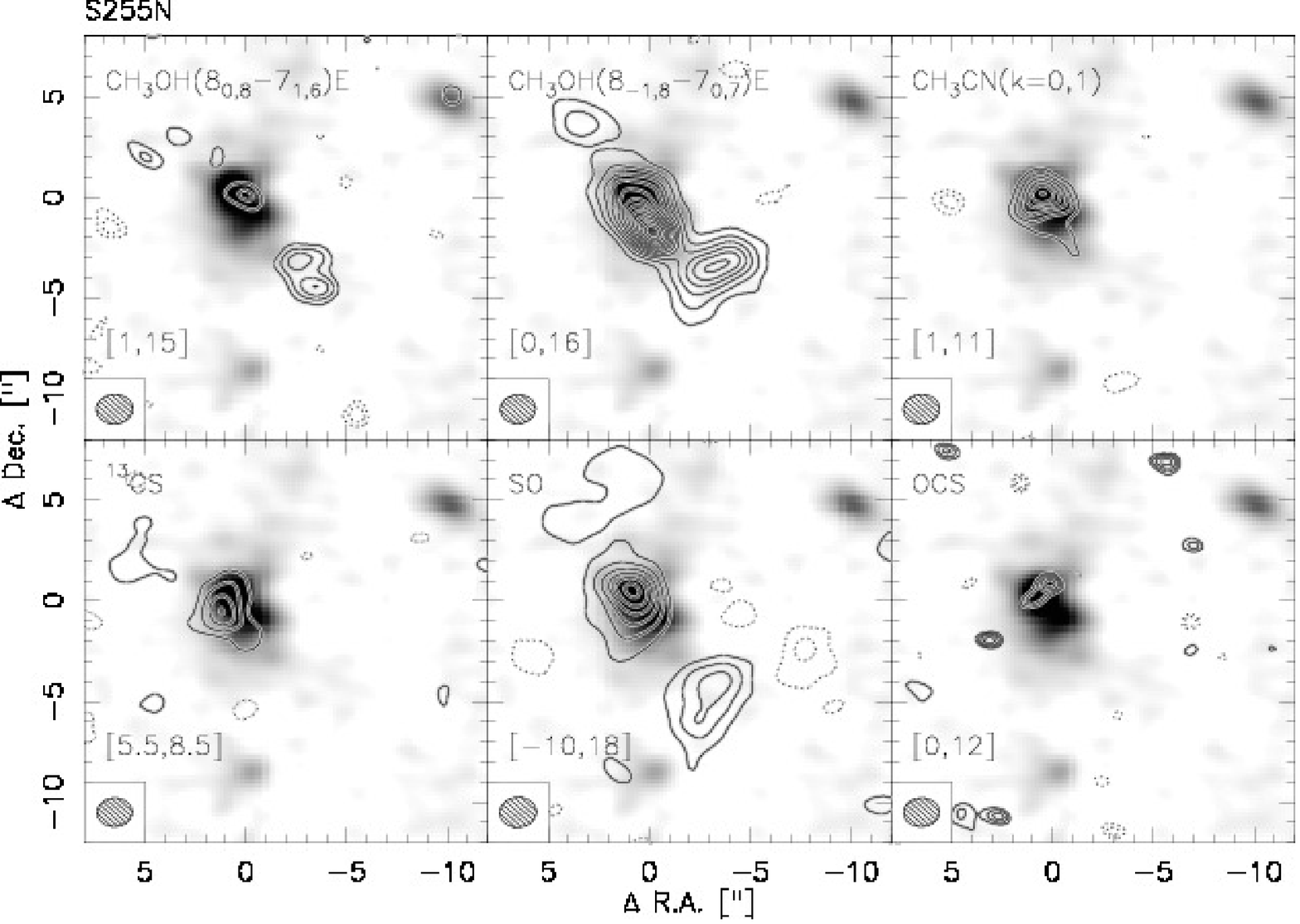}
    \caption{S255N molecular line integrated intensity contour
      with the SMA 1.3 mm continuum emission in the background. 
     All the contour levels start at 3 $\sigma$. The contour step is
     1$\sigma$ in CH$_3$OH($8_{0,8}-7_{1,6}$)A- (1$\sigma=9$ mJy beam$^{-1}$) and CH$_3$CN($k=0\&1$, 1$\sigma=14$ mJy beam$^{-1}$) images, 
     2$\sigma$ in $^{13}$CS (1$\sigma=25$ mJy beam$^{-1}$) image, 3 $\sigma$ in SO (1$\sigma=13$ mJy beam$^{-1}$) and CH$_3$OH($8_{1,8}-7_{0,7}$)E images, and 0.5 $\sigma$ in OCS (1$\sigma=10$ mJy beam$^{-1}$) image. 
     The dotted contours are the negative features. 
     The integral velocity ranges are shown at the bottom-left of each panel in km s$^{-1}$.
    The synthesized beams are shown at the bottom left of each panel. 
    The (0,0) point in each panel is R.A. 06h12m53.669s Dec. $+18^{\circ}00^{\prime}26.90^{\prime\prime}$  (J2000.0).}
\label{nothers}
\end{figure*}

\paragraph{S255N}
In S255N, we detected 10 lines from 6 species ($^{12}$CO, SO, CH$_3$OH, CH$_3$CN, OCS, $^{13}$CS) and 2 additional CO isotopologues ($^{13}$CO and C$^{18}$O) with lower energy levels $E_{lower}/k$ between 5.3 to 65 K (Table \ref{s255nlines}). Figure \ref{nothers} shows the integrated line images of all species (except three CO isotopologues which will be discussed in Sec. \ref{outflow} and \ref{rotation}). All the lines show a peak associated with S255N-SMA1, CH$_3$OH and SO lines also show extended emission along the direction of the outflows (Figure \ref{s255noutcom}, see Sec. \ref{outflow}). 

\paragraph{S255S}
In S255S, we detected 4 lines from two species ($^{12}$CO and SO) and 2 additional CO isotopologues ($^{13}$CO and C$^{18}$O) with lower energy levels $E_{lower}/k$ between 5.3 to 24 K (Table \ref{s255slines}), which again indicates that S255S is an extremely young and cold region. We measure the single dish $^{13}$CO line width toward the S255S SCUBA 850 $\mu$m continuum peak, and calculate the virial mass in this region, which is $\sim$ 80 $M_{\odot}$. The virial mass is smaller than the mass we got from the SCUBA 850 $\mu$m measurement which implies that S255S region will likely collapse.

From the single dish $^{13}$CO observations, we got the v$_{lsr}$ of the three regions, which are shown in Table \ref{vlsr}. However, the  v$_{lsr}$ of the high-mass mm cores in S255IR and S255N based on the interferometer observations is different from the one we got from the single dish data. We extracted dense gas spectra, i.e. CH$_3$OH for S255N-SMA1, CH$_3$CN for S255IR-SMA1 and C$^{18}$O for S255IR-SMA2, and then a Gaussian profile was fitted to get the v$_{lsr}$ of the cores. All the velocities are shown in Table \ref{vlsr}. For S255S we can not get a proper v$_{lsr}$ from the SMA observations due to the missing flux problem, so only the  v$_{lsr}$ we got from 30m observation is listed.

\begin{table}
\caption{The v$_{lsr}$ measured with different telescopes.}
\label{vlsr}
\centering
\renewcommand{\footnoterule}{}  % to avoid a line before footnotes
\begin{tabular}{lcc}
\hline \hline
Sources      &v$_{lsr}$ got from 30m &  v$_{lsr}$ got from SMA\\
~                 &km s$^{-1}$                                    &km s$^{-1}$ \\
\hline
\noalign{\smallskip}
  S255N-SMA1    & 8.6 				& 10.4\\
  S255IR-SMA1   & 7.7 				& 5.2 \\
  S255IR-SMA2   & 7.7				& 9.3\\
  S255S	     &6.6				&...\\
\hline
\end{tabular}
\end{table}

\subsection{The molecular outflows}
\label{outflow}

\begin{figure}[htbp]
   \centering
    \includegraphics[angle=-90,width= 8cm]{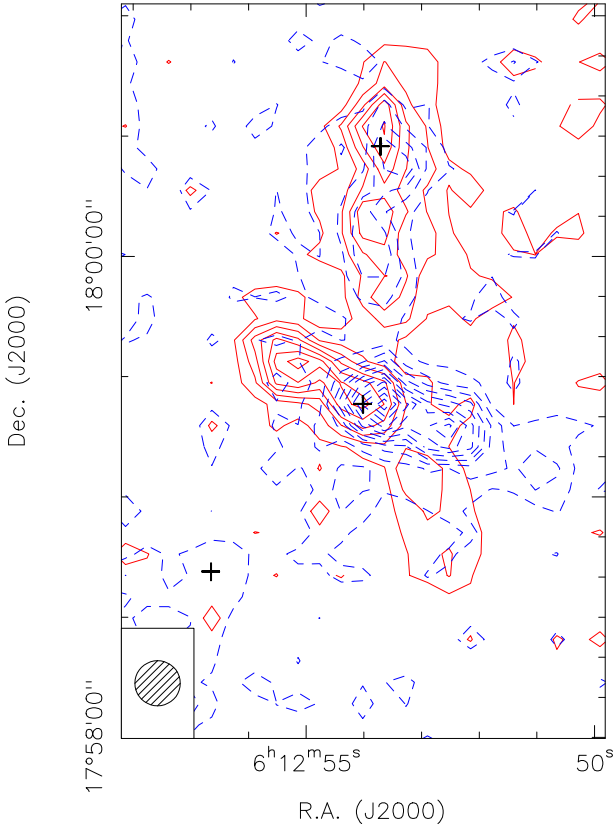}
    \caption{The single dish outflow map of the S255 complex.   The blue-shifted outflow is shown in dashed contours with a velocity-integration regime of [$-$40,0] km s$^{-1}$, and  the red one is shown in full contours with a velocity-integration regime of [16,56] km s$^{-1}$.  The contours start at 4$\sigma$ with a step of 5 $\sigma$($1\sigma$=1.2 K km s$^{-1}$). The three crosses mark the position of the three SCUBA 850 $\mu$m continuum sources.}
\label{s255_out}
\end{figure}

After combining the single dish 30m data with the interferometer data, we integrated the line-wing of $^{12}$CO and produced the outflow image of each region. Figure \ref{s255_out} shows the single dish only outflow map.  While the northeast-southwest outflow which is associated with S255IR is clearly shown in Figure \ref{s255_out}, the high velocity outflow is also found in S255N region, however, the direction of the outflow could not be well defined in single dish outflow map. In the youngest region S255S, the single dish outflow map does not show much emission. We combine the SMA data and 30m data together to study the outflow properties.

\begin{figure*}[htbp]
   \centering
    \includegraphics[angle=-90,width= \textwidth]{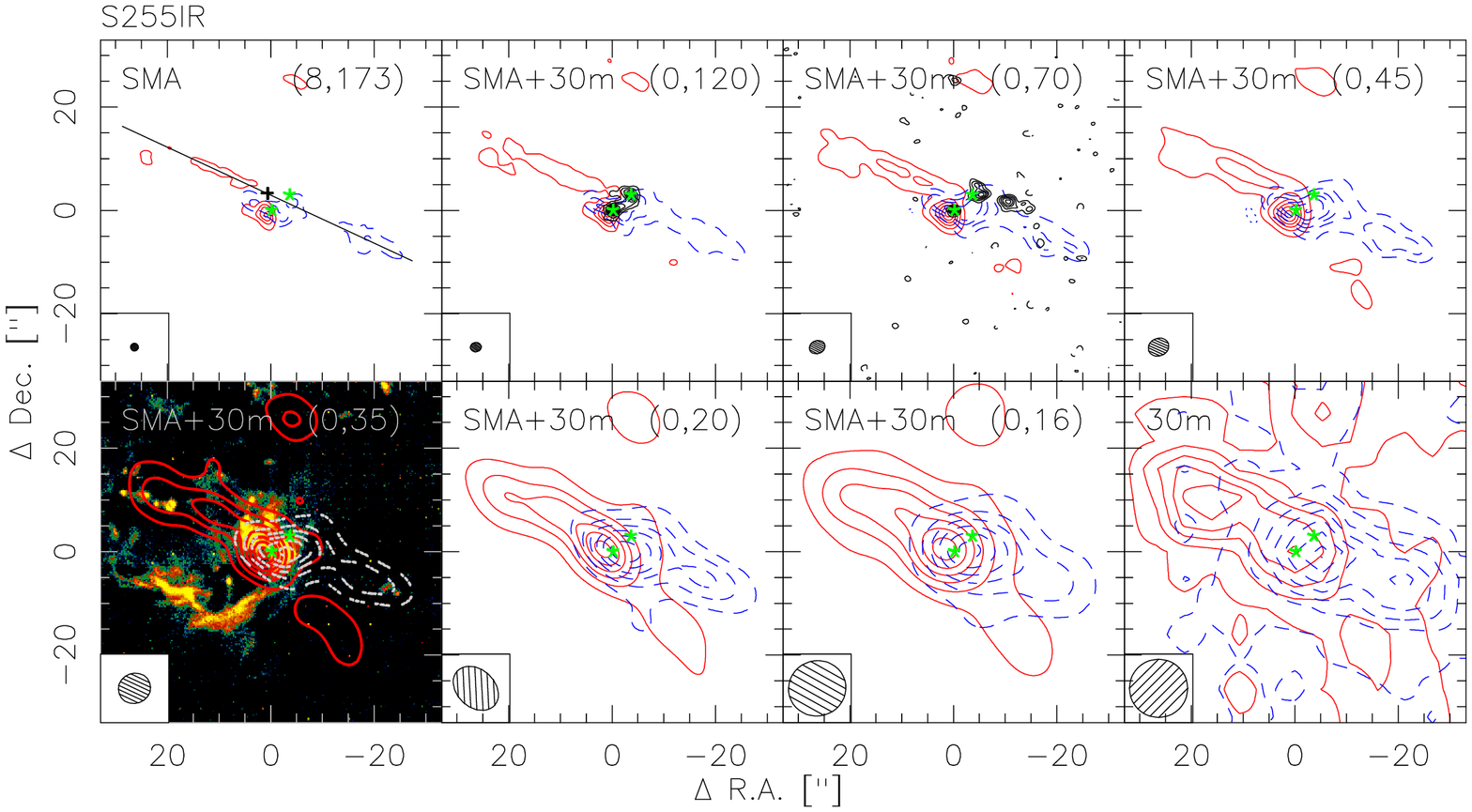}
    \caption{S255IR $^{12}$CO$(2-1)$ outflow images observed with IRAM 30 m telescope and the SMA. 
    The blue-shifted outflow is shown in dashed contours with a velocity-integration regime of [$-$40,0] km s$^{-1}$, and  the red one is shown in full contours with a velocity-integration regime of [16,56] km s$^{-1}$. The top-left panel presents the SMA data only, the bottom-right presents the IRAM 30 m data only. The others are all combined SMA$+$30 m data, which are inverted with different uv-range selections (shown at the top-right of each panel in k$\lambda$) resulting in different beam sizes shown at the bottom left of each panel. All the $^{12}$CO emission contour levels start at 5 $\sigma$ with a step of 10 $\sigma$. The dotted contours are the negative features and the stars mark the position of NIRS 3. In the bottom-left panel, the back ground scale is SINFONI H$_2$ line emission. The line in the SMA-only panel shows the pv-cut presented in Fig. \ref{s255iroutpv}, and the cross marks the central position of the line. In panel (0,120), the SMA 1.3 mm continuum map is over plotted (black full contours), and the contours start at 5 $\sigma$ with a step of 5 $\sigma$. In panel (0,70), the same CH$_3$OH (8$_{-1,8}-7_{0,7}$)E integrated intensity map in Figure \ref{irch3oh} is over plotted in black full contours . The $\sigma$ of the outflow data is shown in Table \ref{outflowdata}. The (0,0) point in each panel is R.A. 06h12m54.019s Dec. $+17^{\circ}59^{\prime}23.10^{\prime\prime}$ (J2000.0).}
\label{s255iroutcom}
\end{figure*}

\paragraph{S255IR}
Figure \ref{s255iroutcom} shows the combined SMA and 30m outflow images of S255IR. The velocity-integration regime of the blue-shifted part is [$-$40,0] km s$^{-1}$ and [16,56] km s$^{-1}$ for red-shifted part. We selected different uv-ranges resulting in different resolutions. Although at different resolution the structure of the outflow changes a lot, the northeast-southwest (NE-SW) outflow is confirmed in all images (also reported by \citet{miralles1997}). In the bottom-left panel, the H$_2$ jet-like sources (region (a) and (b) in Figure \ref{three-color}, see Sec. \ref{linemap}) also follow the NE-SW outflow direction. The water masers which are marked with triangles in the inset of S255IR panel in Figure \ref{continuum} also follow the direction of the outflow, which confirms that S255IR-SMA1 is the driving source of this NE-SW outflow. In all panels except the bottom-right two, the red-shifted part of the outflow bends a little toward northwest, which might be a signature of the precessing jets. In the S255IR panel in Figure \ref{continuum}, the water masers also follow the NE-SW direction. We believe that NIRS 3 is the driving source of this outflow. As we are applying different uv-range selection, two red-shifted outflow components reveal themselves out to the north and south of the continuum sources, respectively, and these two components are both only shown in the lower velocity regime which is offset from the v$_{lsr}$ 10 to 20 km s$^{-1}$. Finally, in the 30m-only panel and Figure \ref{s255_out}, the red-shifted outflow component to the north of the continuum sources becomes connected with the component in S255N.

\begin{figure}[htbp]
   \centering
    \includegraphics[angle=-90,width= 8cm]{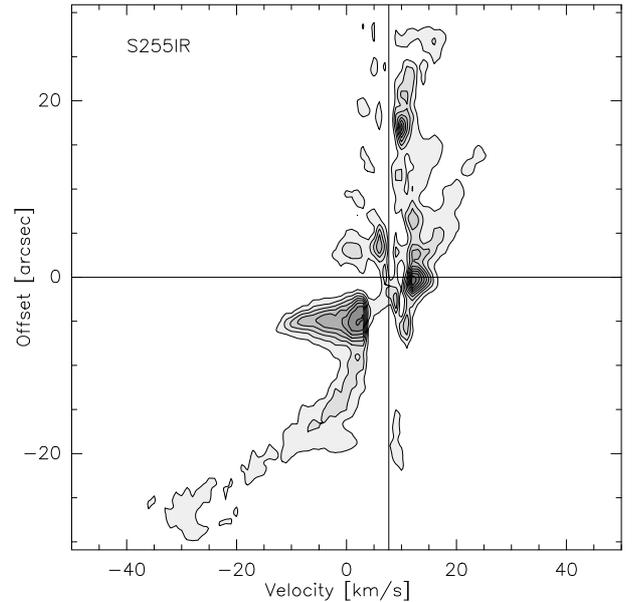}
    \caption{Position-velocity diagrams of S255IR for the $^{12}$CO$(2-1)$ SMA-only outflow observations with a velocity resolution of 1 km s$^{-1}$. The pv-diagram is in northeast-southwest direction with a PA of 75$^{\circ}$ from the north (the cut is  marked in Figure \ref{s255iroutcom}, top-left panel). The contour levels are from 10 to 90\% from the peak emission (4.2 Jy beam$^{-1}$) with a step of 10\%. The $v_{lsr}$ at 7.7 km s$^{-1}$ and the central position (marked by the cross in Fig. \ref{s255iroutcom} top-left panel) is marked by vertical and horizontal line. }
\label{s255iroutpv}
\end{figure}

Figure \ref{s255iroutpv} presents the position-velocity (pv) diagram of the molecular gas along the outflow. The diagram shows that the red- and blue-shifted outflow resembles the Hubble-law with increasing velocity at longer distance from the outflow center, and the blue-shifted outflow also resembles the jet-bow-shock gas entrainment model \citep{arce2007}.

\begin{figure*}[htbp]
   \centering
    \includegraphics[angle=-90,width= \textwidth]{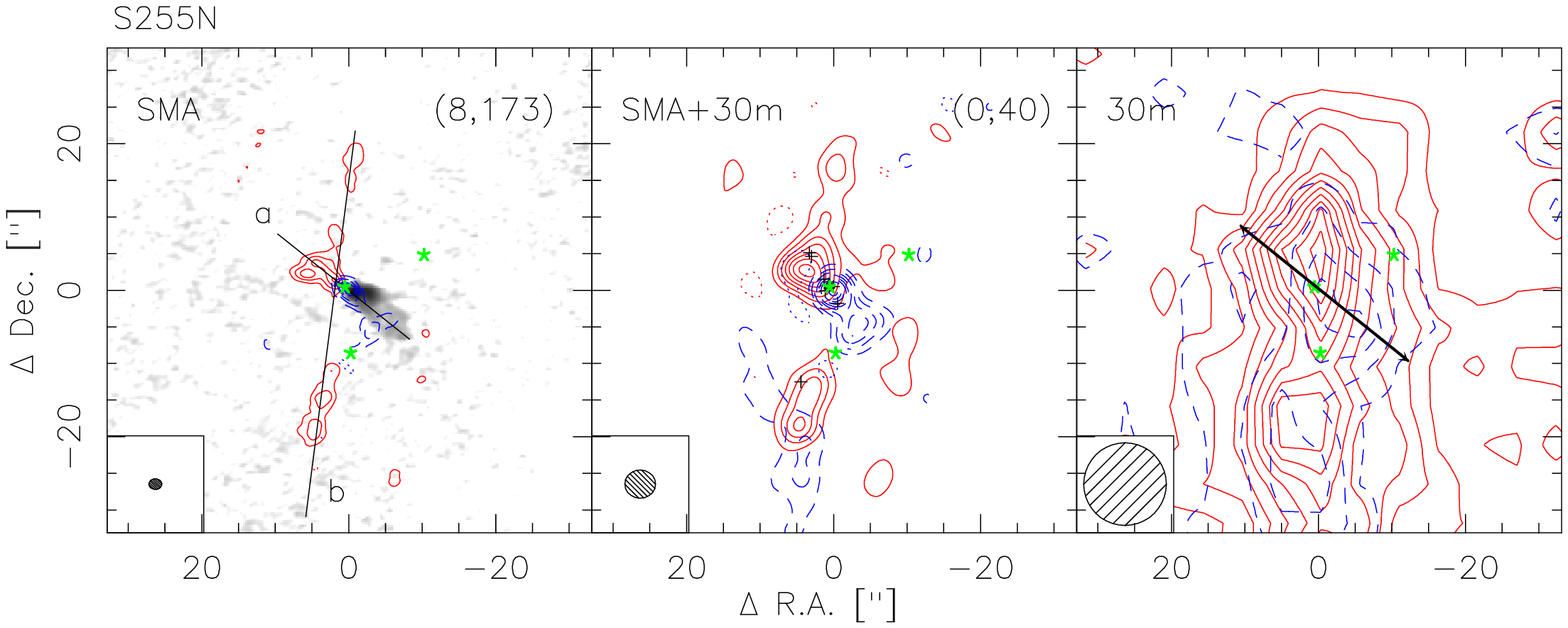}
    \caption{S255N $^{12}$CO$(2-1)$ outflow images observed with IRAM 30 m telescope and the SMA. 
    The blue-shifted outflow is shown in dashed contours with a velocity-integration regime of [$-$44,0] km s$^{-1}$, and the red one is shown in full contours with a velocity-integration regime of [16,48] km s$^{-1}$. The left panel presents the SMA data only, the right presents the IRAM 30 m data only. The middle one is combined SMA$+$30 m data, the uv-range selection is shown at the top-right of each panel in k$\lambda$. All the CO emission contour levels start at 5 $\sigma$ with a step of 6 $\sigma$. The dotted contours are the negative features and the asterisks mark the position of continuum peaks. In the middle panel, the crosses mark the position of the Class I 44 GHz ($7_0-6_1$) A$^+$ methanol masers  detected by \citet{kurtz2004} and the triangle is the water maser \citep{cyganowski2007}. In the top left panel, the back ground grey scale is VLA Q-band continuum \citep{cyganowski2007} and the line shows the pv-cut presented in Fig. \ref{s255noutpv}, which follows the masers direction and crosses the continuum peak. The arrow in the 30m-only panel marks the outflow size we used to calculate the outflow physical parameters. The $\sigma$ of the outflow data is shown in Table \ref{outflowdata}. The (0,0) point in each panel is R.A. 06h12m53.669s Dec. $+18^{\circ}00^{\prime}26.90^{\prime\prime}$  (J2000.0).}
\label{s255noutcom}
\end{figure*}

\paragraph{S255N}
Figure \ref{s255noutcom} shows the combined SMA and 30m outflow images of S255N. The velocity-integration regime of the blue-shifted part is [$-$44,0] km s$^{-1}$ and [16,48] km s$^{-1}$ for the red-shifted part. The northeast-southwest (NE-SW) outflow is shown clearly in the SMA and SMA combined with 30m panels but is hardly seen in the 30m-only panel. In the middle panel, the crosses mark the position of the Class I 44 GHz ($7_0-6_1$) A$^+$ methanol masers  detected by \citet{kurtz2004} and the triangle is the water maser \citep{cyganowski2007}. In the left panel, the back ground grey scale is VLA Q-band 3.6 cm continuum \citep{cyganowski2007} which shows a jet-like structure. All these features follow the NE-SW direction. Based on these features we confirm the direction of the main outflow as NE-SW. The red-shifted outflow to the north of S255N-SMA1 on the line (b) {top left panel in Figure \ref{s255noutcom}} might be part of the outflow cavity, but we can not exclude the possibility of multiple outflows. The red-shifted gas at the bottom part of line (b) seems to be associated with S255N-SMA3 (the middle panel of Figure \ref{s255noutcom}), however, another blue-shifted feature at the southern part of the map shows up as the resolution changes. These two components both can also only be detected at the relatively lower velocity regime $\sim$10 to 20 km s$^{-1}$ offset from the v$_{lsr}$. 

\begin{figure}
   \centering
    \includegraphics[angle=-90,width= 8cm]{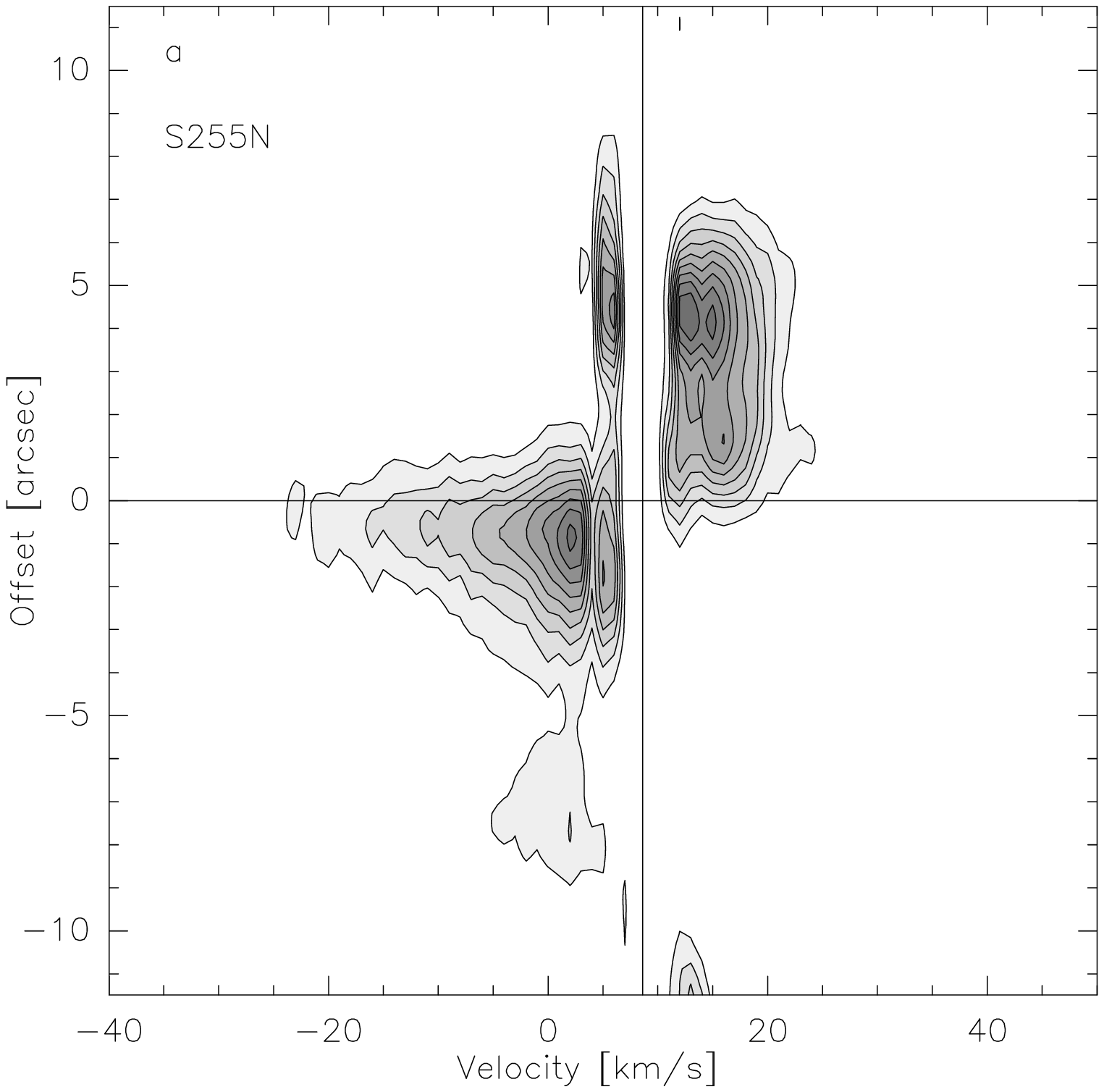}
    \includegraphics[angle=-90,width= 8cm]{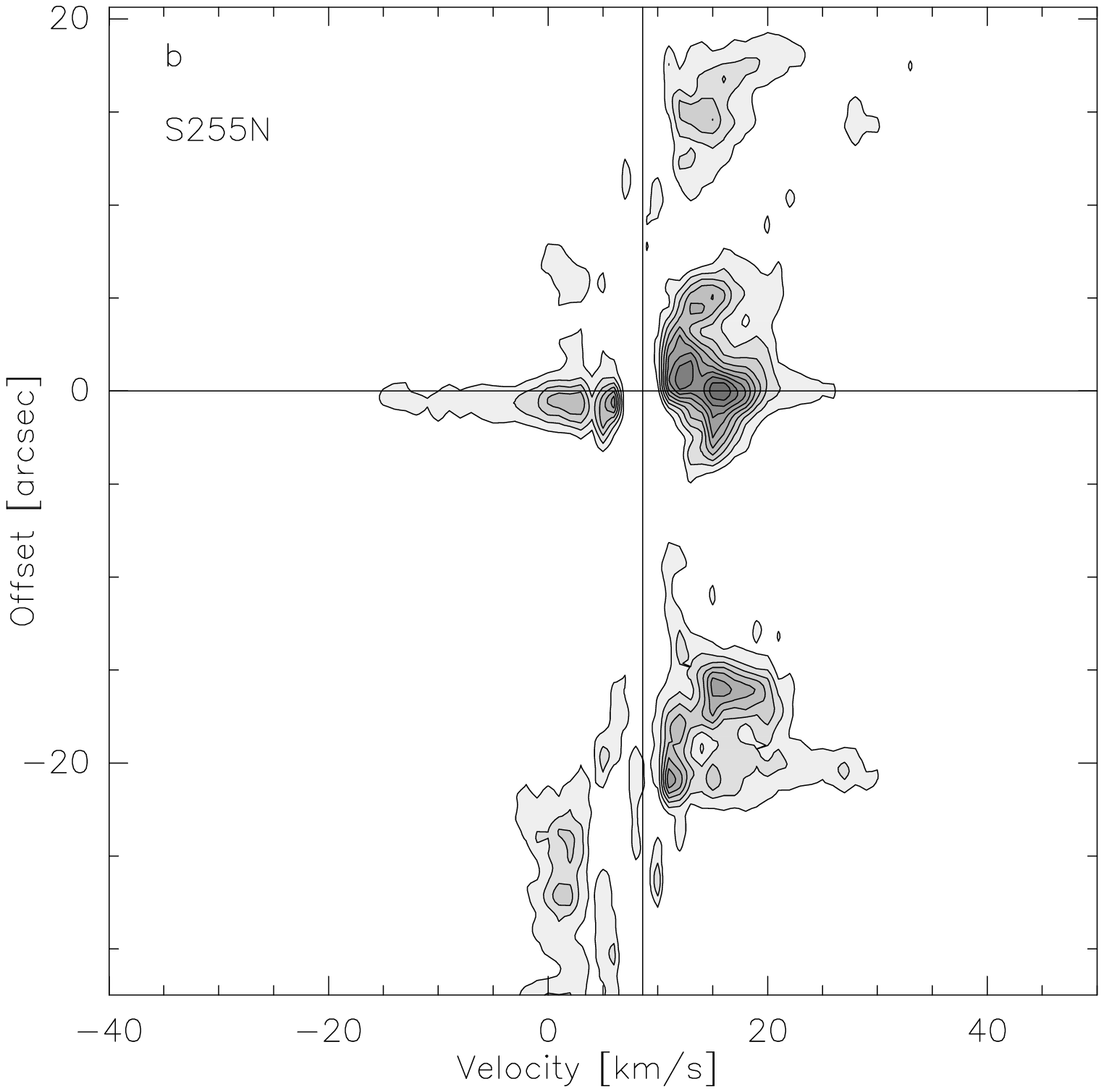}
    \caption{Position-velocity diagrams of S255N for the $^{12}$CO$(2-1)$ SMA-only outflow observations with a velocity resolution of 1 km s$^{-1}$. The top panel is the pv-plot in northeast-southwest direction with a PA of 54$^{\circ}$ from the north (the cut is  marked as the line a in Figure \ref{s255noutcom}, left panel), and the bottom panel is the pv-plot in northwest-southeast direction with a PA of 7$^{\circ}$ from the north (the cut is  marked as the line b in Figure \ref{s255noutcom}, left panel). The contour levels are from 10 to 90\% from the peak emission (4.15 Jy beam$^{-1}$ in the top panel and 2.82 Jy beam$^{-1}$ in the bottom panel) with a step of 10\%. The $v_{lsr}$ at 8.6 km s$^{-1}$ and the central position (i.e. the S255N-SMA1 position in the top panel and the cross point of line a and line b in Figure \ref{s255noutcom} top-left panel in the bottom panel) are marked by horizontal and vertical line.}
\label{s255noutpv}
\end{figure}

Figure \ref{s255noutpv} presents the position-velocity (pv) diagrams of the molecular gas along the outflows. In the top panel, the pv-plot cut follows the direction of the VLA jet-like emission (Fig. \ref{s255noutcom}, left panel, line a). The diagram shows that the high-velocity gas on the blue-shifted side remains very close to the outflow center and the red-shifted part does not show the Hubble-law signature. But in the bottom panel, in which the pv-plot cut follows the direction of the two elongate red-shifted emission (Fig. \ref{s255noutcom}, left panel, line b), the red-shifted northern side of the outflow resembles the Hubble-law. However, the red-shifted southern part of the outflow seems to have nothing to do with our S255N-SMA1 nor S255N-SMA3. The different blue/red pv-characteristics may be explained by the blue $^{12}$CO emission tracing predominantly the jet like component also visible in centimeter continuum emission, whereas the red $^{12}$CO emission traces mainly the cavity-like walls of the outflow or another outflow.

\begin{figure*}[htbp]
   \centering
    \includegraphics[angle=-90,width= \textwidth]{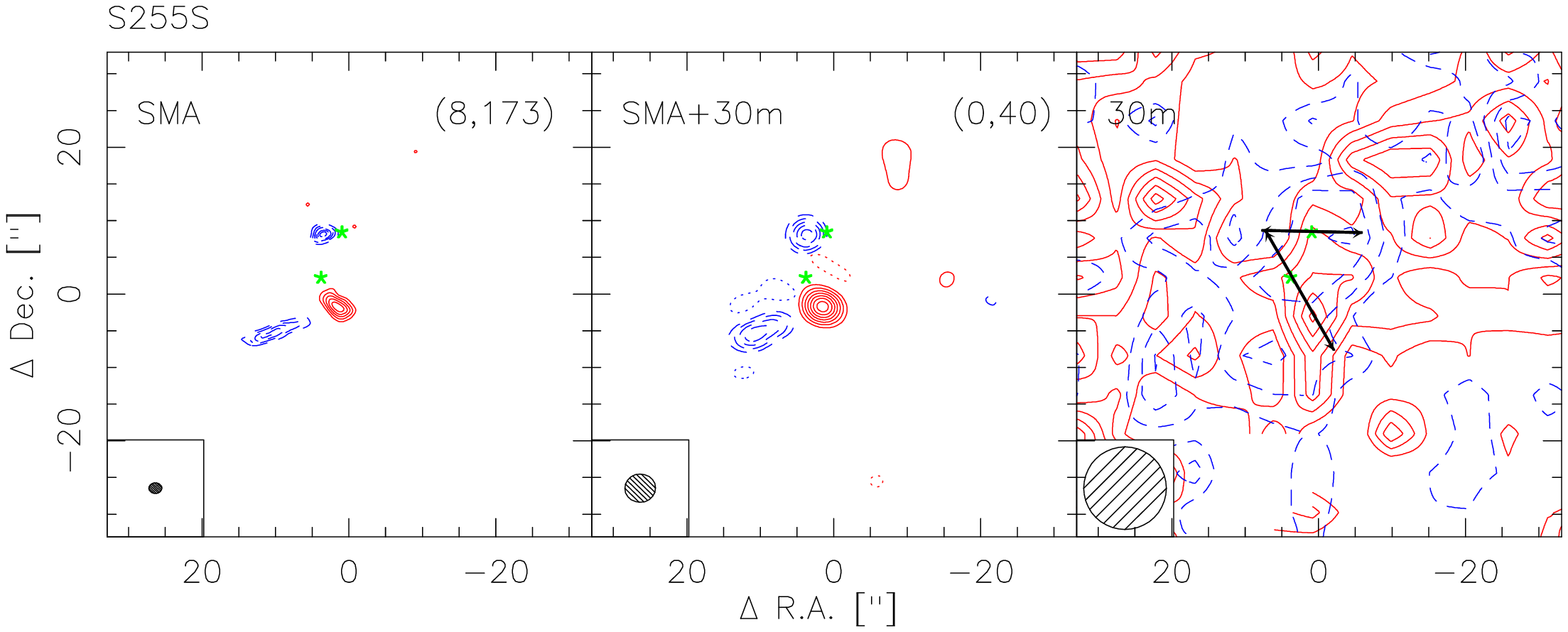}
    \caption{S255S $^{12}$CO$(2-1)$ outflow images observed with IRAM 30 m telescope and the SMA. 
    The blue-shifted outflow is shown in dashed contours with a velocity-integration regime of [$-$9,1] km s$^{-1}$, and the red one is shown in full contours with a velocity-integration regime of [14,32] km s$^{-1}$. The left panel presents the SMA data only, the right presents the IRAM 30 m data only.  The middle one is combined SMA$+$30 m data, and the uv-range selection is shown at the top-right of each panel in k$\lambda$. All the $^{12}$CO$(2-1)$ emission contour levels start at 5 $\sigma$ with a step of 4 $\sigma$, except the one in IRAM 30 m only image which starts at 5 $\sigma$ with a step of 2 $\sigma$. The dotted contours are the negative features and the stars mark the positions of continuum peak S255S-SMA1 and S255S-SMA2. The arrows in the 30m-only panel mark the outflow sizes we used to calculate the outflow physical parameters. The $\sigma$ of the outflow data is shown in Table \ref{outflowdata}. The (0,0) point is each panel corresponds to position R.A. 06h12m56.58s Dec. $+17^{\circ}58^{\prime}32.80^{\prime\prime}$ (J2000.0).}
\label{s255soutcom}
\end{figure*}

\paragraph{S255S} 
Figure \ref{s255soutcom} presents the combined SMA and 30m outflow images of S255S. The velocity-integration regime of the blue-shifted part is [$-9$,1] km s$^{-1}$ and [14,32] km s$^{-1}$ for the red-shifted part. All the panels except the 30m-only one, show two blue-shifted components and one red-shifted component. Although the line wing emission is less pronounced than that for the other two regions, we clearly identify blue- and red-shifted $^{12}$CO emission associated with both mm peaks. The blue-shifted part in the south of the map is only detected in the 1 km s$^{-1}$$\geq$v$\geq-6$ km s$^{-1}$, and the red-shifted part in the north of the map is only detected in the v$\leq$16 km s$^{-1}$. Since the line wing emission does not extend to very high velocities, the outflow should not be oriented directly along the line of sight. Therefore, the small spatial extent of the outflow indicates as well the youth of this region. 

\begin{table*}
\caption{Outflow data.}
\label{outflowdata}
\centering
\renewcommand{\footnoterule}{}  % to avoid a line before footnotes
\begin{tabular}{lccccr}
\hline \hline
\noalign{\smallskip}
Region&uv-range        &beam size               &$\sigma_{\rm red}$&$\sigma_{\rm blue}$&Data\\
\hline
\noalign{\smallskip}
~	   &[k$\lambda$]&[$^{\prime\prime}$]&[mJy beam$^{-1}$]&[mJy beam$^{-1}$]    & ~\\
\hline
\noalign{\smallskip} 
S255IR &(8, 173)     	   &1.5$\times$1.5         &15		 	       &26 	                                           &SMA\\
~	    &(0, 120)	  	  &2.1$\times$1.8	    &15			       &26		                           &SMA+30m\\
~	    & (0, 70)     	  &3.1$\times$2.5	    &18			        &32				    &SMA+30m\\
~	    &(0, 45)		  &4.0$\times$3.4	    &22			        &38				    &SMA+30m\\
~	    &(0, 35)		  &6.3$\times$5.8	    &26			        &71				    &SMA+30m\\
~	    &(0, 20)		  &9.8$\times$7.4	    &38			      &86				    &SMA+30m\\
~	    &(0, 16)		  &11.1$\times$10.7 &40			      &110			                &SMA+30m\\
~	    &...			  &11.3$\times$11.3   &0.8 K km s$^{-1}$         &1.1 K km s$^{-1}$                 &30m\\
S255N&(8, 173)           	 &1.8$\times$1.4           &19			       &12                                               &SMA\\
~	    &(0, 40)		  &4.2$\times$3.9	    &34			       &18				    &SMA+30m\\
~	    &...			  &11.3$\times$11.3   &0.5 K km s$^{-1}$        &1.1 K km s$^{-1}$             &30m\\
S255S&(8, 173)           	  &1.8$\times$1.4           &14		 	      &18                                                &SMA\\
~	    &(0, 40)		  &4.2$\times$3.9	    &20			      &26			   	    &SMA+30m\\
~	    &...			  &11.3$\times$11.3   &0.4 K km s$^{-1}$       &0.5 K km s$^{-1}$   		 &30m\\
\hline
\end{tabular}
\end{table*}

Applying the method of \citet{cabrit1990, cabrit1992} we derived properties of the outflows such as outflow mass, outflow energy and dynamical age. Since the SMA data suffer a lot from the missing flux problem, only the 30m data are used in the calculations. These calculations assume that the $^{13}$CO$(2-1)$/$^{12}$CO$(2-1)$ line wing ratio is 0.1\citep{choi1993, levreault1988, beuther2002b}. The results are shown in Table \ref{outflow_p}. The dynamical age calculation depends on the tangent of the outflow's inclination angle (with 0$^{\circ}$ being in the plane of the sky). Since we cannot get the inclination angle information of these sources, we assumed an inclination angle of 45$^{\circ}$ for all the dynamical age calculations. For S255N and S255S, the outflows are not clearly defined in the 30m-only maps (Figures \ref{s255noutcom} and \ref{s255soutcom}). Therefore, we used with the SMA+30m map to determine the outflow sizes (see right panels of Figures \ref{s255noutcom} and \ref{s255soutcom}). For S255S, since we just detected only one outflow lobe associated with each continuum source, we just extended the only one lobe to the opposite side of the source to get the proper size and mass of the outflow. The arrows in the 30m-only panels in Figures \ref{s255noutcom} and \ref{s255soutcom} mark the outflow size we use.

\begin{table*}
\caption{Outflow parameters.}
\label{outflow_p}
\centering
\begin{tabular}{lccccccccr}
\hline \hline
\noalign{\smallskip}
Sources&$M_{\rm t}$		     &$p$				&$E$			&Size	    &t	             &$\dot{M}_{\rm out}$	            &$F_{\rm m}$\\
\hline
\noalign{\smallskip}
~&[$M_{\odot}$]&[$M_{\odot}$ km s$^{-1}$]			&[erg]     		  &[pc]	&[yr]&[$M_{\odot}$/yr]  &[$M_{\odot}$/km/s/yr] \\
\hline
\noalign{\smallskip} 
  S255IR-SMA1	&2.9		    &137				&6.5$\times10^{46}$&0.7&7.4$\times10^3$&4$\times10^{-4}$&2$\times10^{-2}$\\
  S255N-SMA1	&1.0		    &43				&1.9$\times10^{46}$&0.2&2.4$\times10^3$&4$\times10^{-4}$&2$\times10^{-2}$\\
  S255S-SMA1	&0.02		    &0.4				&8$\times10^{43}$&0.1&2.5$\times10^3$&8$\times10^{-6}$&2$\times10^{-4}$\\
  S255S-SMA2            &0.03		   &0.8				&2$\times10^{44}$&0.07&1.7$\times10^3$&2$\times10^{-5}$&5$\times10^{-4}$\\
\hline
\end{tabular}
\footnotesize{~\\
Entries include total outflow mass $M_{\rm t}$, momentum $p$, energy $E$, size, outflow dynamical age $t$, outflow rate $\dot{M}_{\rm out}$, mechanical force $F_{\rm m}$.}
\end{table*}

\subsection{Rotational structures}
\label{rotation}
\begin{figure*}[htbp]
   \centering
    \includegraphics[angle=-90,width= \hsize]{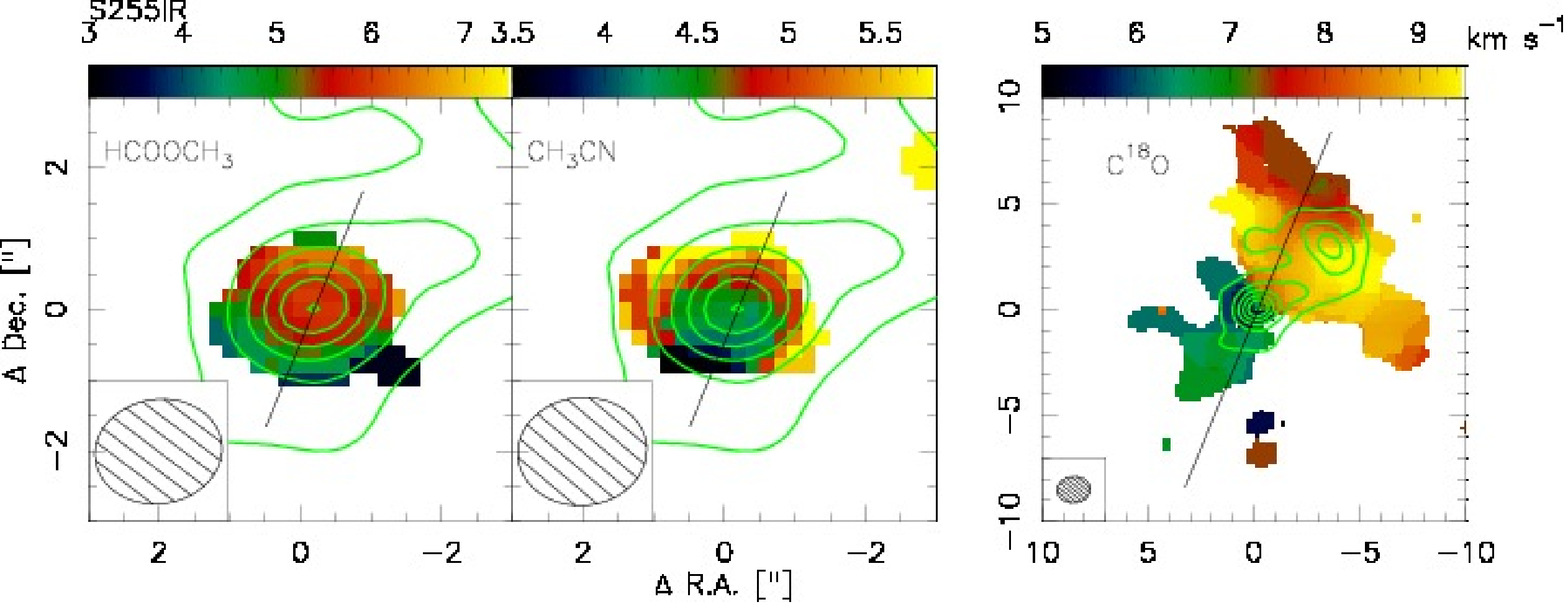}
    \caption{HCOOCH$_3$,  CH$_3$CN ($k=2$) and C$^{18}$O$(2-1)$ velocity (1st) moment maps overlaid with the SMA 1.3 mm dust continuum of S255IR. The contours start at 5$\sigma$ and increase with a step of 10$\sigma$ in all panels ($\sigma$=1.7 mJy beam$^{-1}$). The lines in each panel show the pv-diagram cut presented in Fig. \ref{s255ir_mom1_pv}. All moment maps were clipped at the five sigma level of the respective line  channel map. The synthesized beam is shown in the lower left corner of each plot. The (0,0) point in each panel is R.A. 06h12m54.019s Dec. $+17^{\circ}59^{\prime}23.10^{\prime\prime}$ (J2000.0).}
\label{s255ir_mom1}
\end{figure*}

\paragraph{S255IR}
Figure \ref{s255ir_mom1} shows the velocity moment maps of HCOOCH$_3$, CH$_3$CN ($k=2$) and C$^{18}$O$(2-1)$.  In the velocity maps of HCOOCH$_3$ and CH$_3$CN ($k=2$), we see a clear velocity gradient perpendicular to the outflow axis, which indicates the existence of a rotational structure. The C$^{18}$O$(2-1)$ velocity map shows a little different picture compared to the other ones. C$^{18}$O traces a more diffuse gas compared to the other two molecules. The C$^{18}$O velocity map also shows a big velocity difference between S255IR-SMA1 and S255IR-SMA2. 

\begin{figure*}[htbp]
   \centering
    \includegraphics[angle=-90,width= 14cm]{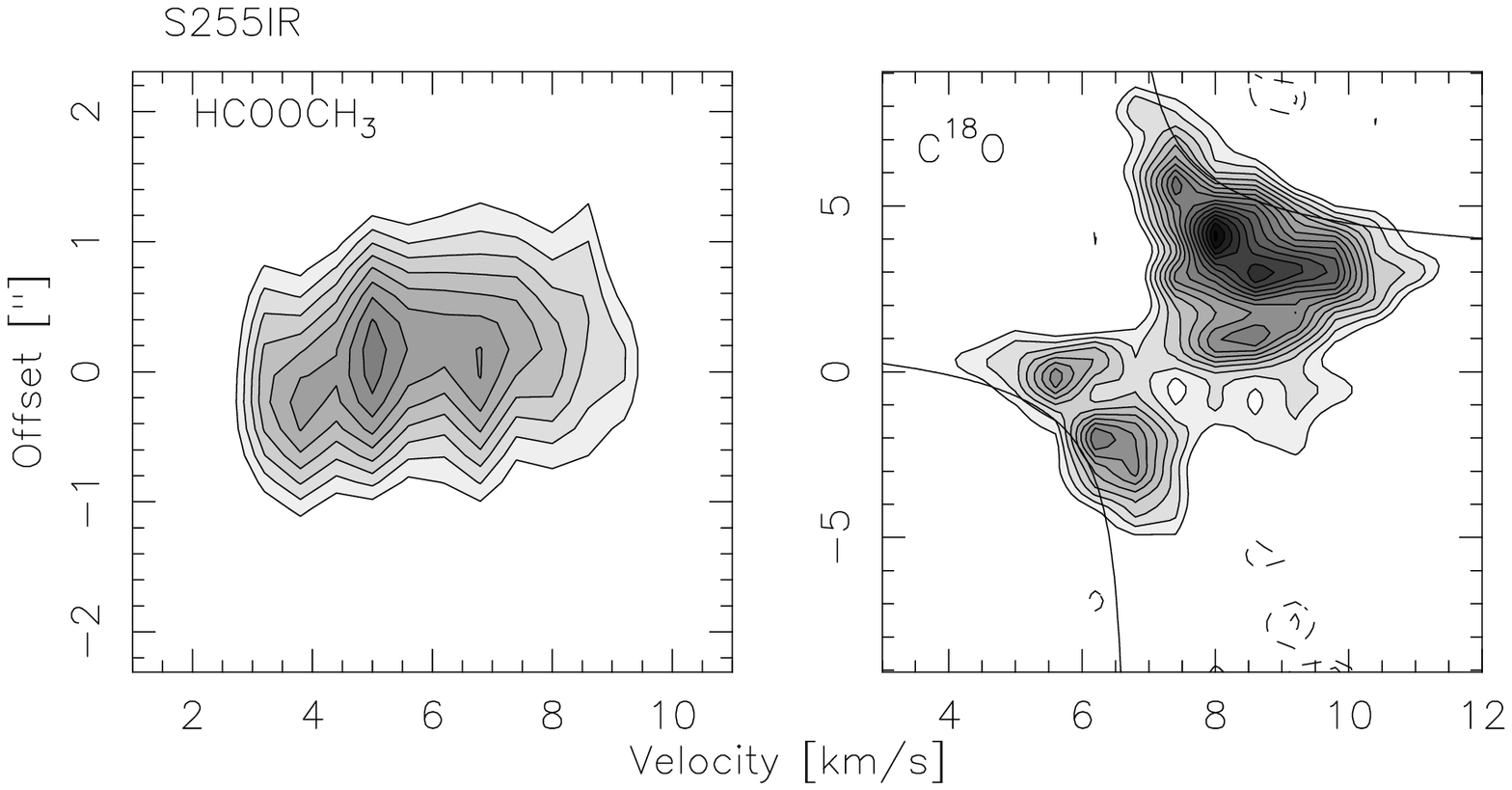}
    \caption{Position-velocity diagrams derived for the cuts along the observed velocity gradient in Fig. \ref{s255ir_mom1}. The offset refers to the distance along the cut from the dust continuum peak. The contour levels are from 3$\sigma$ with a step of 1$\sigma$ in both panels (1$\sigma$=60 mJy beam$^{-1}$). The full line in the C$^{18}$O panel shows a Keplerian rotation curve with a central mass of 28 $M_{\odot}$, which is the mass of the whole continuum structure in S255IR. The negative features are shown in dashed lines.} 
\label{s255ir_mom1_pv}
\end{figure*}

Figure \ref{s255ir_mom1_pv} shows the position-velocity diagrams of the HCOOCH$_3$ and C$^{18}$O$(2-1)$ emission. The cuts, which are shown in Figure \ref{s255ir_mom1}, go through the peak of the dust continuum and have position angles perpendicular to the direction of the outflow presented in Figure \ref{s255iroutcom}. The pv-diagram  of HCOOCH$_3$ shows that the rotational structure is not in Keplerian motion, hence maybe it is just a rotating and infalling core similar to the toroids described by \citet{cesaroni2007}. The approximately 2.3$^{\prime\prime}$ diameter of the rotational structure corresponds to 3 700 AU at the given distance of 1.59 kpc. The pv-diagram of CH$_3$CN ($k=2$) has a similar structure to the one of HCOOCH$_3$, thus we do not show it. 

However, the picture is different for C$^{18}$O, the pv-diagram shows a Keplerian-like rotation structure. The full line in the C$^{18}$O panel shows a Keplerian rotation curve with a central mass of 28 $M_{\odot}$, which is the mass of the whole continuum structure in S255IR covering both mm peaks. For a more detailed discussion, see Sec. \ref{disk_can}.

\begin{figure*}[htbp]
   \centering
    \includegraphics[angle=-90,width= \hsize]{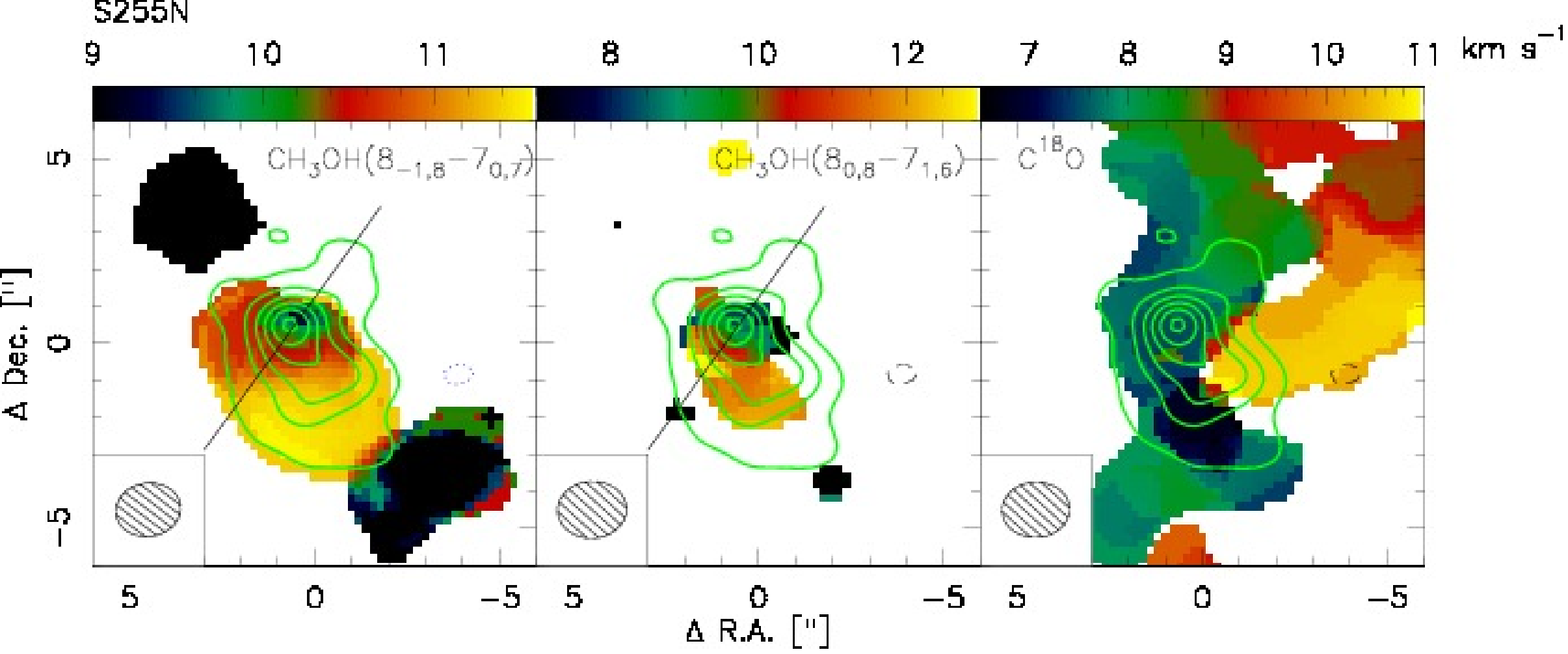}
    \caption{CH$_3$OH(8$_{-1,8}-7_{0,7}$),  CH$_3$OH ($8_{0,8}-7_{1,6}$) and C$^{18}$O$(2-1)$ velocity (1st) moment maps overlaid with the SMA 1.3 mm dust continuum of S255N. The contours start at 5$\sigma$ and increase with a step of 10$\sigma$ (1$\sigma$=1.6 mJy beam$^{-1}$). All moment maps were clipped at the five sigma level of the respective line's channel map. The lines in each panel show the pv-diagram cut presented in Fig. \ref{s255ir_mom1_pv}. The synthesized beam is shown in the lower left corner of each plot. The (0,0) point in each panel is R.A. 06h12m53.669s Dec. $+18^{\circ}00^{\prime}26.90^{\prime\prime}$  (J2000.0).}
\label{s255n_mom1}
\end{figure*}

\paragraph{S255N}
Figure \ref{s255n_mom1} shows the velocity moment maps of CH$_3$OH(8$_{-1,8}-7_{0,7}$),  CH$_3$OH ($8_{0,8}-7_{1,6}$) and C$^{18}$O$(2-1)$. The C$^{18}$O velocity map shows a complicated velocity structure, the velocity gradient is neither aligned with nor orthogonal to the outflow orientation. Because C$^{18}$O is an isotopologue of $^{12}$CO, it may be influenced by infall and the outflow. Methanol is well known as a molecule that traces cores, shocks and masers in star formation regions \citep{jorgensen2004, beuther2005a, sobolev2007}. It has also been reported as a low mass disk tracer \citep{goldsmith1999}. In the velocity maps of the two methanol transitions (left and middle panel of Figure \ref{s255n_mom1}), we see a velocity gradient perpendicular to the outflow axis, which indicates the existence of a rotational structure. 

\begin{figure}[htbp]
   \centering
    \includegraphics[angle=-90,width= 8cm]{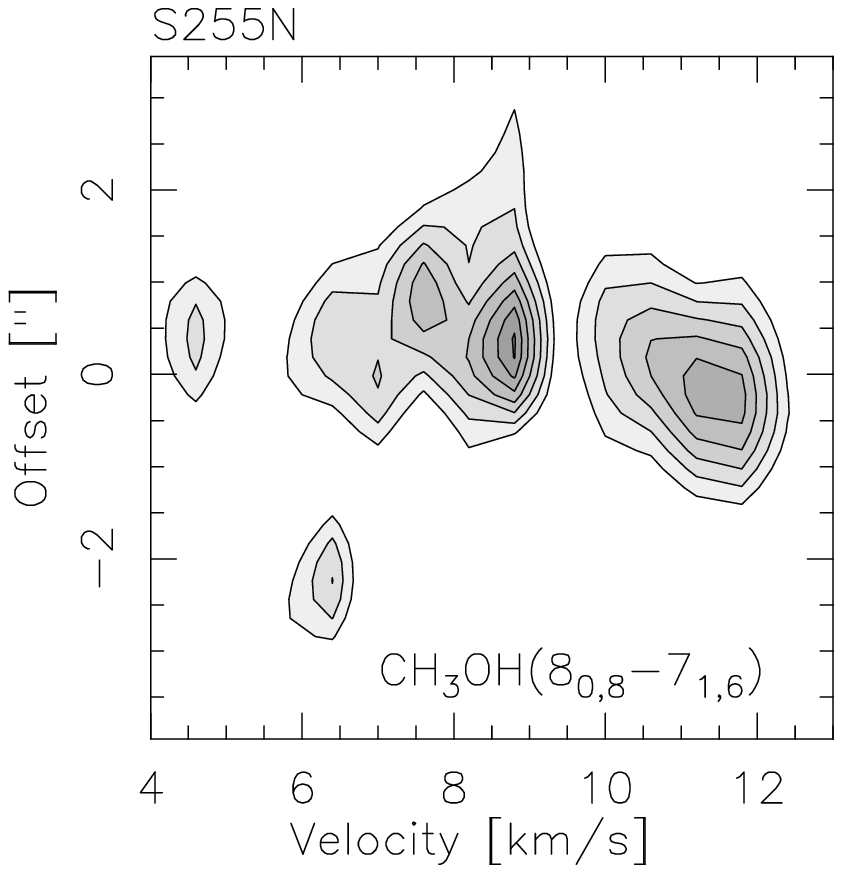}
    \caption{Position-velocity diagrams derived for the cuts along the observed velocity gradient in Fig. \ref{s255n_mom1}. The offset refers to the distance along the cut from the dust continuum peak S255N-SMA1. The contour levels are all from 3$\sigma$ and with a step of 1$\sigma$ (1$\sigma$=40 mJy beam$^{-1}$).}
\label{s255n_mom1_pv}
\end{figure}

Figure \ref{s255n_mom1_pv} shows the position-velocity diagram of the (8$_{0,8}-7_{1,6}$) methanol line emission, and the pv-diagram of the other methanol transition shows the similar velocity structure. The cuts go through the peak of the dust continuum and have position angles perpendicular to the direction of the outflow (Figure \ref{s255n_mom1}). The pv-diagram shows that the rotational structure is not Keplerian. The structure has a size of 4.4$^{\prime\prime}$ corresponding to 7 000 AU at the given distance of 1.59 kpc. It has a narrow velocity range of 3 km s$^{-1}$, and may likely be a large rotating and infalling core similar to the toroids described by \citet{cesaroni2007}.

There are two low-velocity components at the northeast and southwest of the continuum peak in the methanol velocity maps which coincide with the outflow. They may be caused by the shock heating emission, and the line profile is consist with both thermal emission and maser emission  \citep{sutton2004, sobolev2005, kalenskii2002, slysh2002}.

\subsection{Temperature from CH$_3$CN(12$_k-11_k$) in S255IR}
\label{ch3cn_tem}
\begin{figure}[htbp]
   \centering
    \includegraphics[angle=-90,width= 8cm]{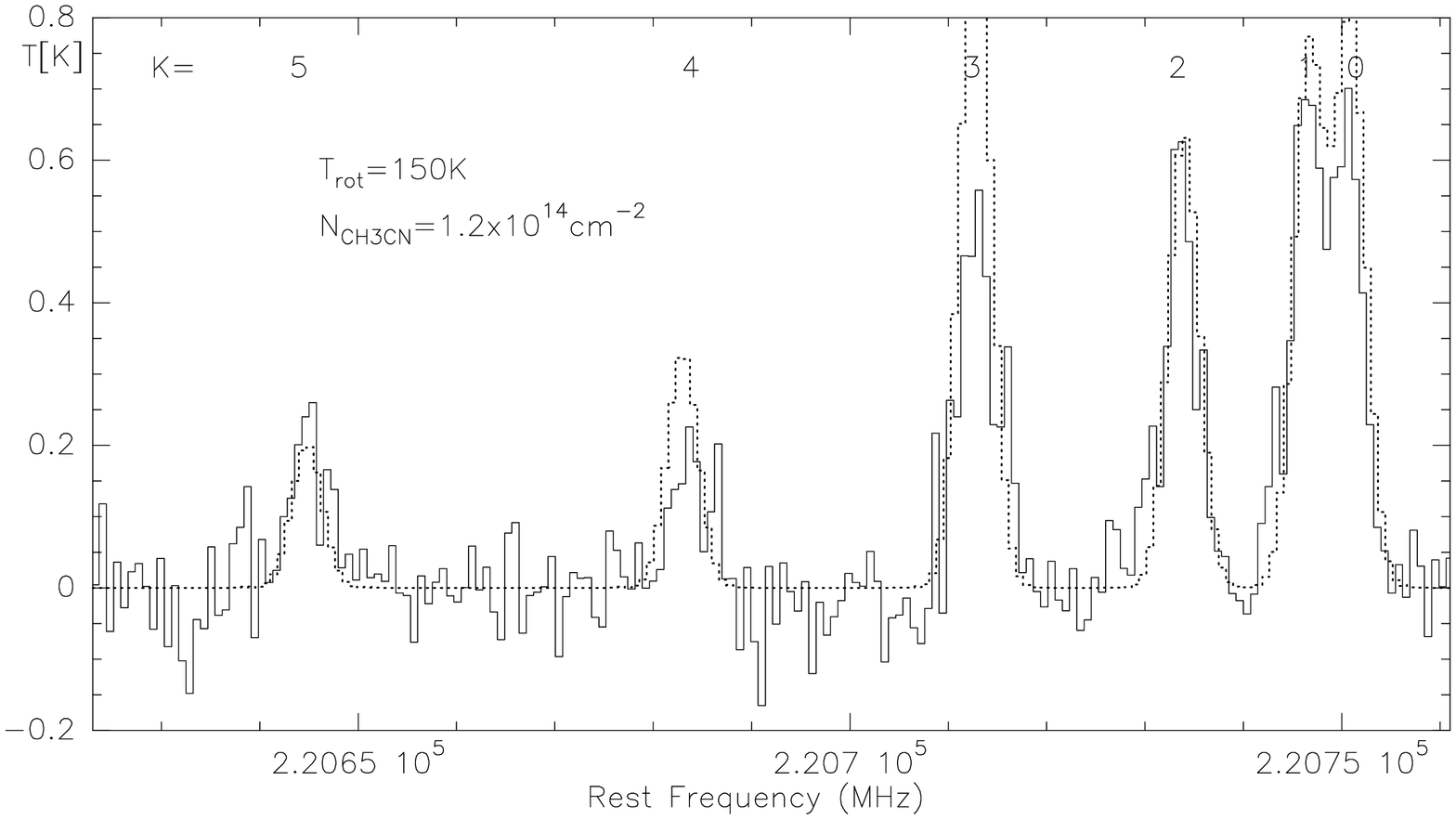}
    \caption{CH$_3$CN(12$_k $- 11$_k$) spectrum toward the mm continuum peak S255IR-SMA1. The dotted line shows a model spectrum with $T_{\rm rot}$=150 K and $N_{\rm CH_3CN}=3.5\times10^{14}$ cm$^{-2}$. The $k=6$ line is excluded in the  model because it is blurred by the HNCO(10$_{1,9}-9_{1,8}$) line.}
\label{s255irtemper}
\end{figure}

\begin{figure}[htbp]
   \centering
    \includegraphics[angle=0,width= 8cm]{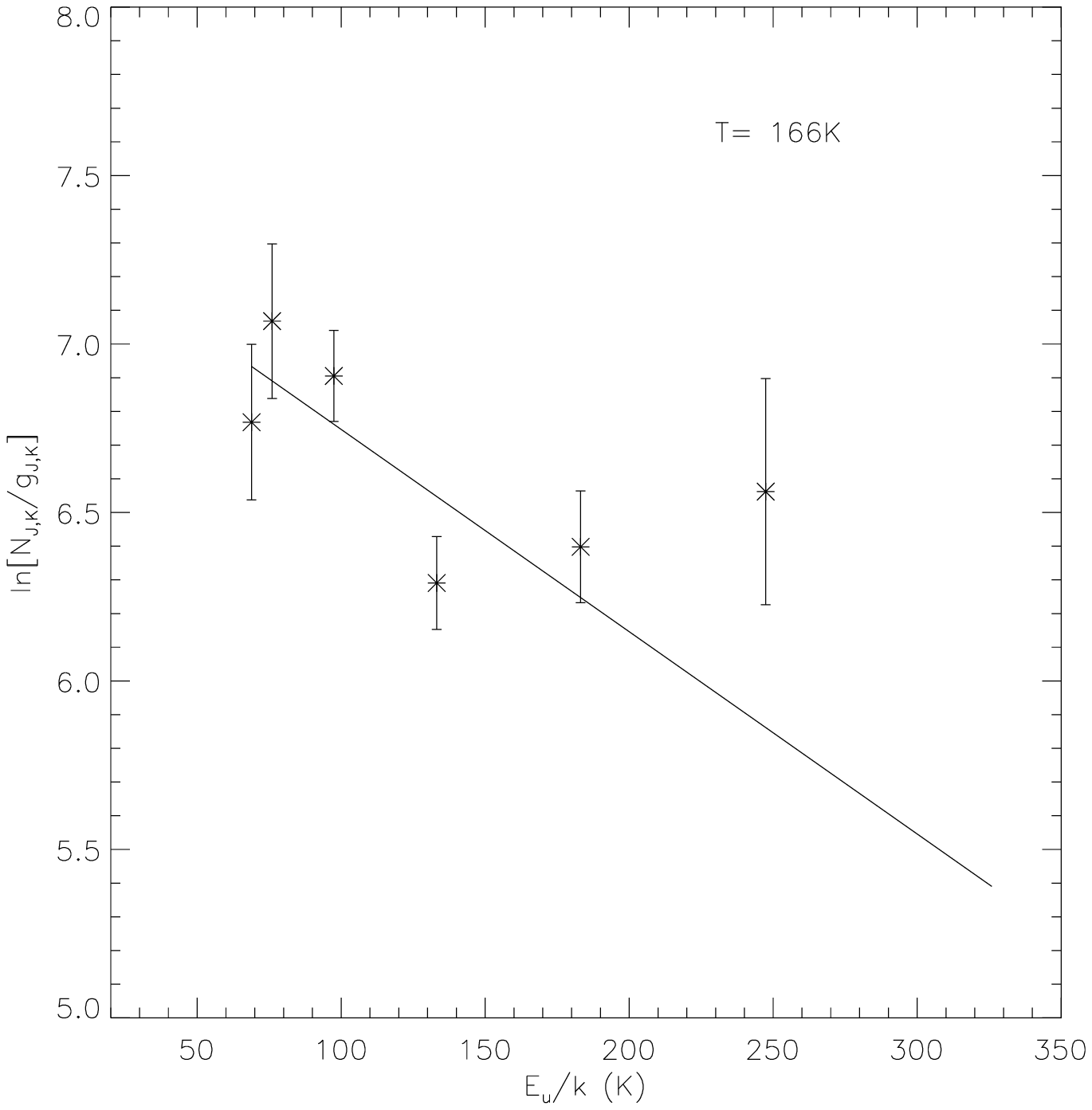}
    \caption{From the Gaussian fitting result, we calculated the $N_{j,k}$ \citep{zhang1998}, and plotted in the figure above. $E_u$ is the upper level energy of each transition. The line shows the linear fitting of the first five points.}
\label{s255irtemperfit}
\end{figure}

Since we detected 7 lines of the CH${_3}$CN$(12_k-11_k)$ $k$-ladder with $k$ = 0...6 in S255IR, we can utilize the varying excitation levels of the lines with lower level energies $E_{lower}/k$ between 58 to 315 K (Table \ref{s255irlines}) to estimate a temperature for the central gas core. Figure \ref{s255irtemper} shows the observed CH${_3}$CN$(12_k-11_k)$ spectrum toward the continuum peak  S255IR-SMA1 with only the compact configuration data. We did a simple Gaussian fitting of the spectrum and plot the level populations $N_{j,k}$ we calculated from the fitting result in Figure \ref{s255irtemperfit} with the assumption of optically thin emission \citep{zhang1998}. The linear fitting result of the lower five levels is also plotted in Figure \ref{s255irtemperfit}. It is clear that we can not fit the whole spectrum with one single temperature, which reveals a temperature gradient of the source and not optical thin emission. We modeled this spectrum in the local thermodynamic equilibrium using the XCLASS software developed by Peter Schilke (private communication). This software package uses the line catalogs from JPL and CDMS \citep{poynter1985,mueller2001}. The model spectrum with a temperature of 150 K is shown in Figure \ref{s255irtemper} in dotted line. The main difference between the model spectrum and the observed spectrum is that the model one is optically thin whereas the lower line intensity of the observed $k=3$ line indicates moderate optical depth of the CH$_3$CN lines. 

\subsection{SINFONI results}

\subsubsection{SINFONI line and maps}
\label{linemap}
\begin{figure}[htbp]
   \centering
    \includegraphics[angle=0,width= 8cm]{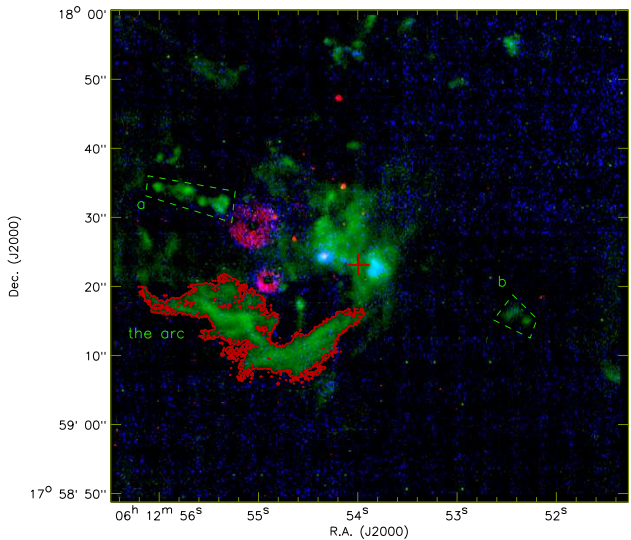}
    \caption{Three-color image created from SINFONI line maps. Red: Br$\gamma$ line emission, green: H$_2$ (2.12 $\mu$m) line emission and blue: Fe {\tiny II} (1.64 $\mu$m) line emission. The cross marks the position of S255IR-SMA1. The contour and the dashed lines mark the regions of which the excitation diagrams are constructed to identify the nature of the emission (Fig. \ref{nebula_temp}). }
\label{three-color}
\end{figure}

The SINFONI observations are centered on the mm peak S255IR-SMA1 covering most part of the near-infrared cluster. Figure \ref{three-color} shows the 3 color composite of 3 line maps (Red: Br$\gamma$ line emission, green: H$_2$ (2.12 $\mu$m) line emission and blue: Fe {\tiny II} (1.64 $\mu$m) line emission). The two massive young stars (NIRS 1 and NIRS 3) are in the center of the field, surrounded with the cluster of about 120 stars down to $K=$17 mag. For all the point sources marked in Figure \ref{position} a SINFONI $H$- and $K$-band spectrum is available and a spectral classification is obtained of  the brighter members (Sec. \ref{classification}). The PDR and jet like emission is traced by the molecular H$_2$ emission (2.12 $\mu$m, green). And to the north of the S255IR-SMA1 (the asterisk in the Figure \ref{three-color}), two point sources (sources \#17 and \#18 in Figure \ref{position}) show strong Br$\gamma$, which may indicate the existence of accreting signature \citep{muzerolle1998}. The two strong Fe {\tiny II} (blue) features around S255IR-SMA1 show destructive shock (J-type, \citet{hollenhach1989}) emission and also follow the direction of the outflow and the H$_2$ jets, which indicate S255IR-SMA1 is the energetic driving source of the jets. 

\begin{figure}[htbp]
   \centering
    \includegraphics[angle=90,width= 8cm]{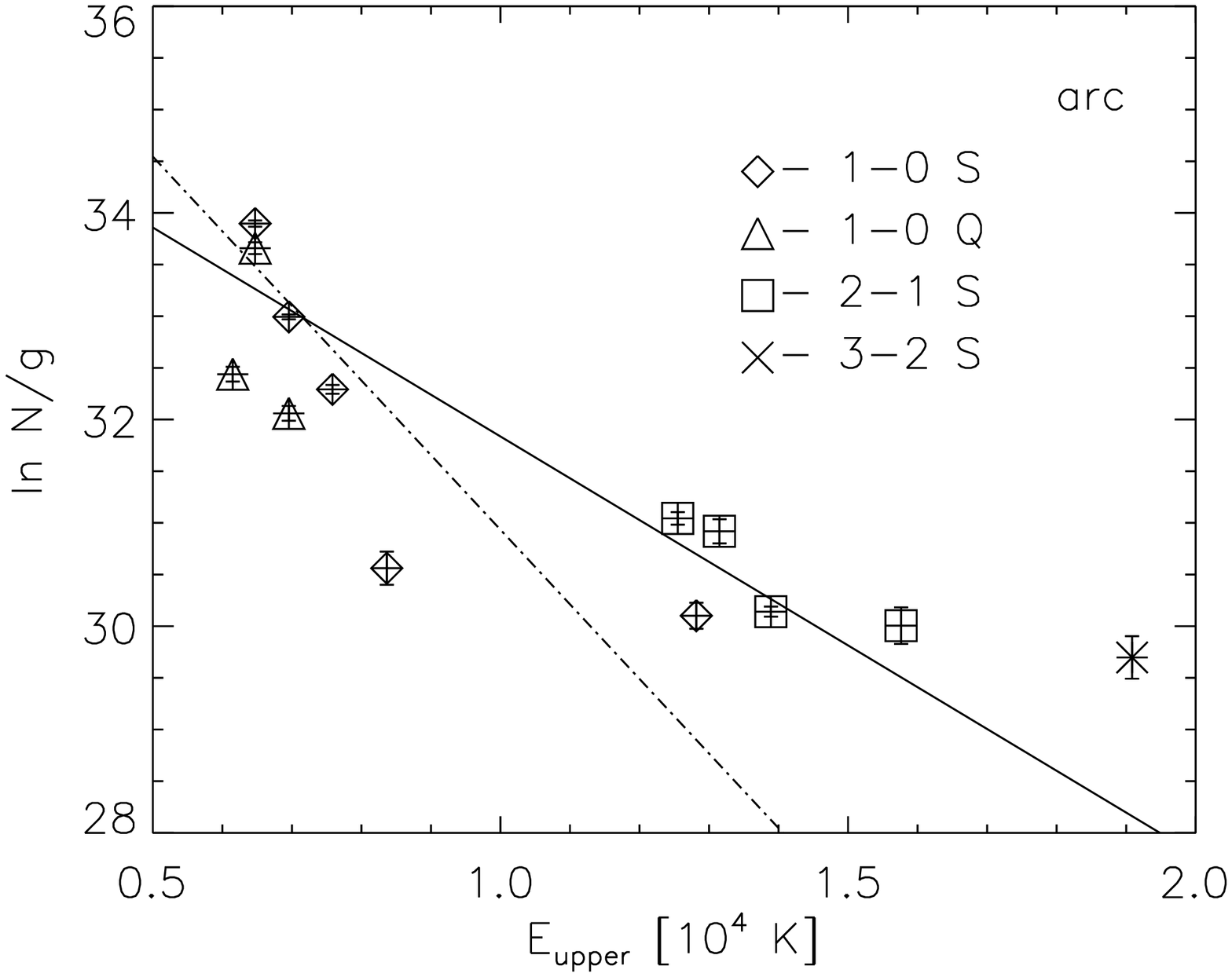}
    \includegraphics[angle=90,width= 8cm]{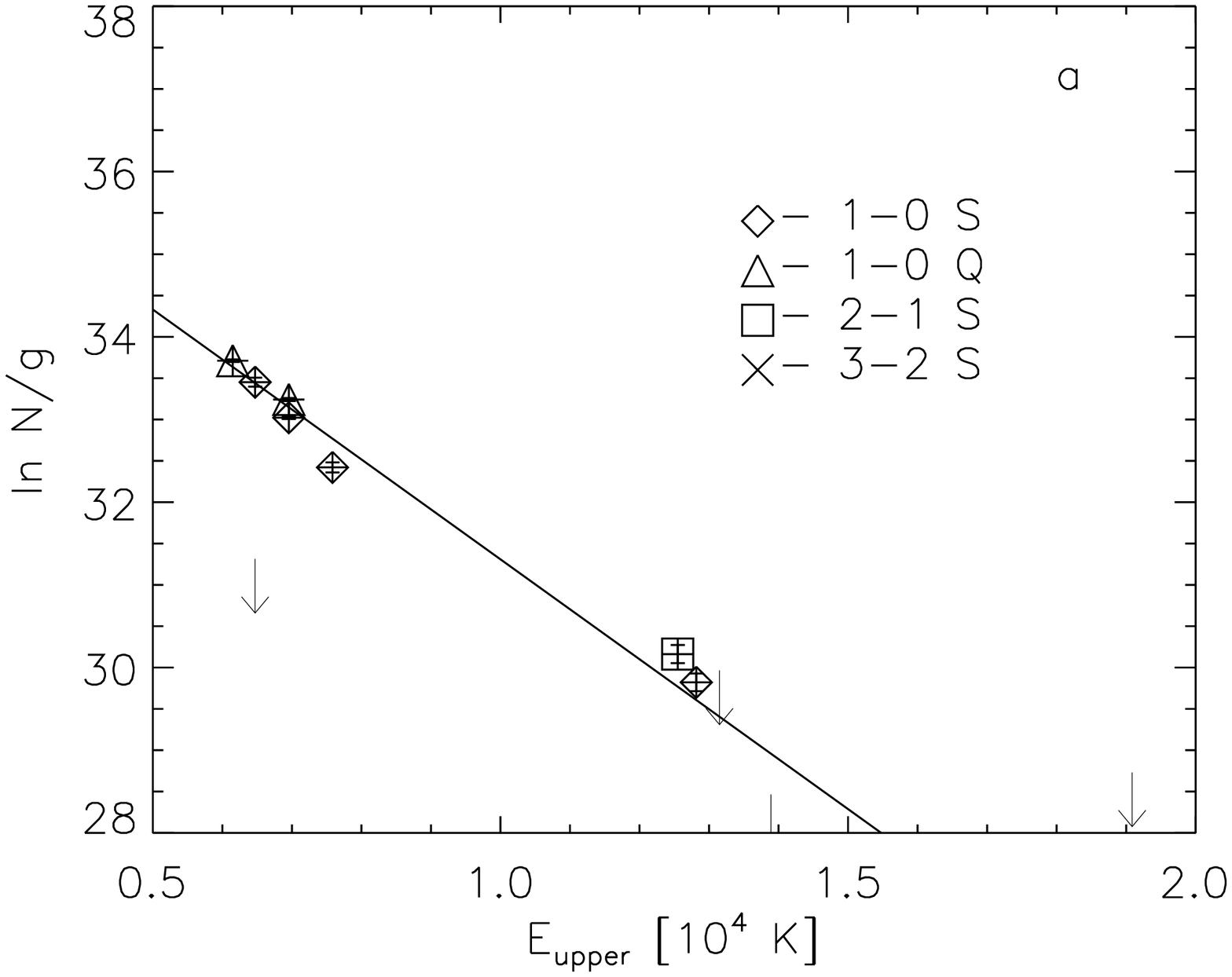}
    \includegraphics[angle=90,width= 8cm]{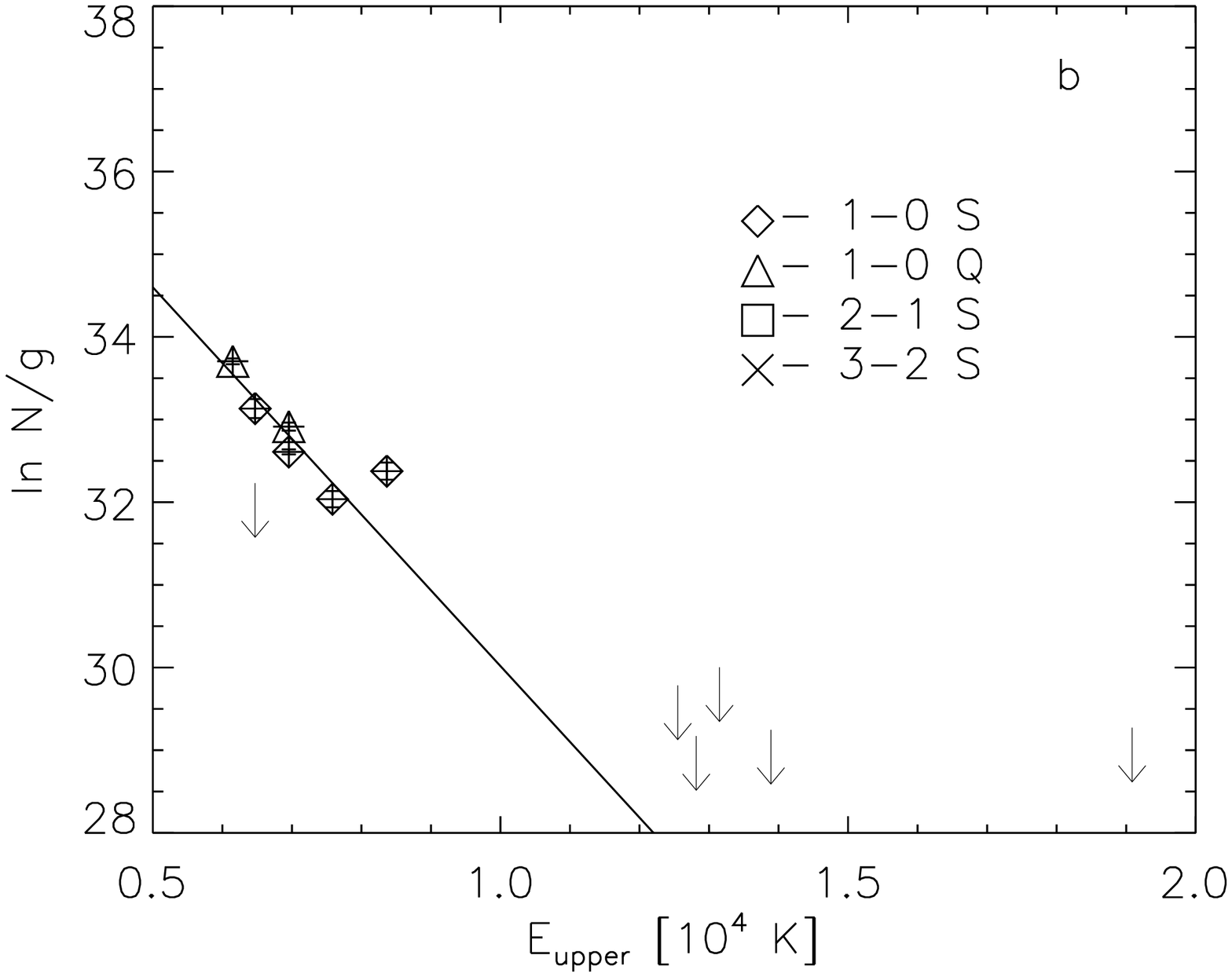}
    \caption{The excitation diagram of the whole arc in the south-east of the Fig. \ref{three-color} (the top panel) and two jet-like emissions (a) and (b). The solid line in all diagrams is a single temperature fit to all the lines detected, while for the dash-dotted line in the top panel a only the 1$-$0 S lines are included in the fit. The excitation diagram of the arc suggests that UV fluorescence is the excitation mechanism in this area. Regions (a) and (b) are most likely outflows as their excitation diagrams suggest thermal excitation.}
\label{nebula_temp}
\end{figure}

To identify the excitation mechanism of the H$_2$ emission (the arc which is marked in Figure \ref{three-color} by the red contour and the jet like emission regions  (a) and (b)), we extracted the spectra of these three regions and construct the excitation diagrams for these regions. Figure \ref{nebula_temp} shows the excitation diagrams of them. In these diagrams, the measured column densities of lines are plotted against the energy of the upper level (see \citet{martin-hernandez2008} for the detailed description of this diagram). The total column densities were calculated using the description of \citet{zhang1998}. Different symbols represent different vibrational levels (Fig. \ref{nebula_temp}). The arc region (Fig. \ref{three-color}) has the most lines detected as it covers the largest area. For the jet-like emission region (a) and (b), the weaker lines seen in the spectrum of the arc region are not detected and 3$\sigma$ upper limits are given instead. The solid line in the diagrams is a single temperature fit to all the data points, while the dashed line in the diagram of the arc region presents a linear fit to only the $1-0$ S lines. For the arc region it is clear that the column densities are not represented by a single temperature gas, the line fluxes of the $2-1$ and $3-2$ vibrational levels are higher than expected from the $1-0$ line fluxes. A likely excitation mechanism of the H$_2$ gas in the arc is fluorescence by non-ionizing UV photons \citep{davis2003}. 

The excitation diagrams of regions (a) and (b) are well-represented by a single temperature. However the $3-2$ lines are not detected in either region, and only one $2-1$ line is detected in region (a) left most of them only upper limits. We conclude that the excitation diagrams of both regions are consistent with shock excited emission in an outflow emission. Besides, the bottom left panel of Figure \ref{s255iroutcom} shows region (a) and (b) follow the direction of the outflow, also their elongated shapes that all suggest outflow origin. The linear fits of the excitation diagrams also allow us to determine the column density and temperature of the emitting gas (Table \ref{h2gas}).

\begin{table}
\caption{Physical properties of the H$_2$ gas.}
\label{h2gas}
\centering
\renewcommand{\footnoterule}{}  % to avoid a line before footnotes
\begin{tabular}{lcc}
\hline \hline
Region      & T$_{rot}$     &  N(H$_2$)\\
~                &$[K]$    & [cm$^{-2}$] \\
\hline
\noalign{\smallskip}
  the arc  &$2416\pm64$   & $2.6\pm0.2\times10^{17}$ \\
   a           &$1655\pm62$   & $5.8\pm1.7\times10^{17}$\\
   b          & $1091\pm91$   & $1.9\pm1.2\times10^{18}$\\
\hline
\end{tabular}
\end{table}

\subsubsection{SINFONI stellar spectral type classification}
\label{classification}
Our SINFONI observations provide an $H$- and $K$-band spectrum of each source with a spectral resolution of R=1500. SOFI $J$- and $K$-band photometry results are taken from \citet{bikthesis}. Objects with $K<$14 have high  enough S/N spectra to obtain a reliable spectral type. This reduces the sample to 39 sources, however, we can only get the spectral type of 16 sources which are listed in Table \ref{spectype}. For many point sources, we could not get a spectral type, some of them have very flat spectra and basically no absorption lines and also no infrared excess, they are likely to be gas clumps. Others have very red spectra, which are likely dominated by dust emission from the surrounding environment or circumstellar disks, therefore, the spectrum of the underlying objects might not be visible, such as sources \#17, \#18 in Table \ref{spectype} and our infrared sources NIRS1 and NIRS3 which will be discussed later. 

\begin{figure*}[htbp]
   \centering
    \includegraphics[angle=0,width= \textwidth]{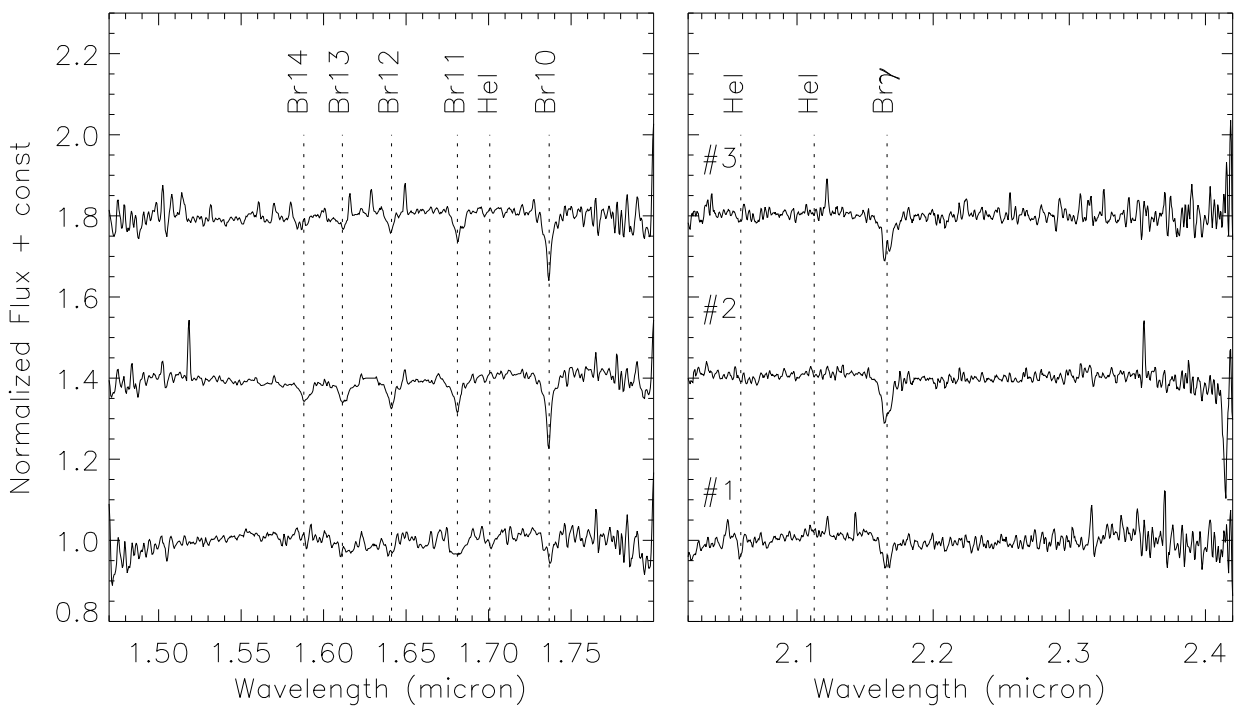}
    \caption{Normalized SINFONI $H$-band (left panel) and $K$-band (right panel) spectra of the three B/A stars detected in S255IR. Their spectra are characterized by absorption lines of the hydrogen Bracket series and He {\tiny{I}} (dotted vertical lines). The stars are numbered according to Table \ref{spectype}.}
\label{obspec}
\end{figure*}

The SINFONI $H+K$-band spectra of the three brightest cluster members (sources \#1, \#2 and \#3) show Br$\gamma$ absorption in the $K$-band and Br$10-14$ absorption in the $H$-band (Figure \ref{obspec}). Weaker lines of He {\tiny I} are also visible in $H$ (1.70 $\mu$m) and K (2.058 $\mu$m).  We apply the $K$-band classification scheme for B stars from \citet{hanson1996} which links a $K$-band spectral type to an optical spectral type. For the $H$-band, we used the classification of \citet{hanson1998} and \citet{blum1997}. Table \ref{ewofearly} shows the measured Equivalent Widths (EW) of the relevant lines in the $H$- and $K$-band spectra of the three B/A stars.  Star \#1 has a strong Br$\gamma$ absorption but does not show He {\tiny{I}} (2.11 $\mu$m), therefore is classified as B3V$-$B7V. Star \#2 and \#3 show much stronger Br$\gamma$ absorption and also do not have He {\tiny{I}} (2.11 $\mu$m) are therefore classified as B8V$-$A3V. The EWs of the $H$-band lines are in agreement with the $K$-band spectral type. Source \#1 and source \#2 are also associated with UCH{\tiny II} regions \citep{snell1986}, which suggests that they are high to intermedia mass stars, and this consists with our spectral type classification results. The results are listed in Table \ref{spectype}.

\begin{table*}
\caption{Equivalent width and $K$-band spectral types of the early type stars in S255IR.}
\label{ewofearly}
\centering
\renewcommand{\footnoterule}{}  % to avoid a line before footnotes
\begin{tabular}{lrrrrrr}
\hline \hline
Star& Br11 (1.68 $\mu$m)&He {\tiny{I}} (1.70 $\mu$m)& He {\tiny{I}} (2.11$\mu$m)&Br$\gamma$ (2.166 $\mu$m)&$K$-Spectral type&Optical Spectral type\\
\hline
\noalign{\smallskip}
  1  &$<$0.4 \AA  		      &0.6$\pm0.4$ \AA			&$<$0.4 \AA		&5.9$\pm0.4$ \AA			&KB4$-$B7	&B3V$-$B7V\\
  2   &8.3$\pm0.4$ \AA	     &$<$0.4 \AA				&$<$0.4 \AA		&11$\pm0.4$ \AA			&KB8$-$A3	&B8V$-$A3V\\
  3   &8.9$\pm0.4$ \AA	    &$<$0.4 \AA				&$<$0.4 \AA		&8.9$\pm0.4$ \AA			&KB8$-$A3	&B8V$-$A3V\\
\hline
\end{tabular}
\end{table*}

\begin{figure*}[htbp]
   \centering
    \includegraphics[angle=0,width= \textwidth]{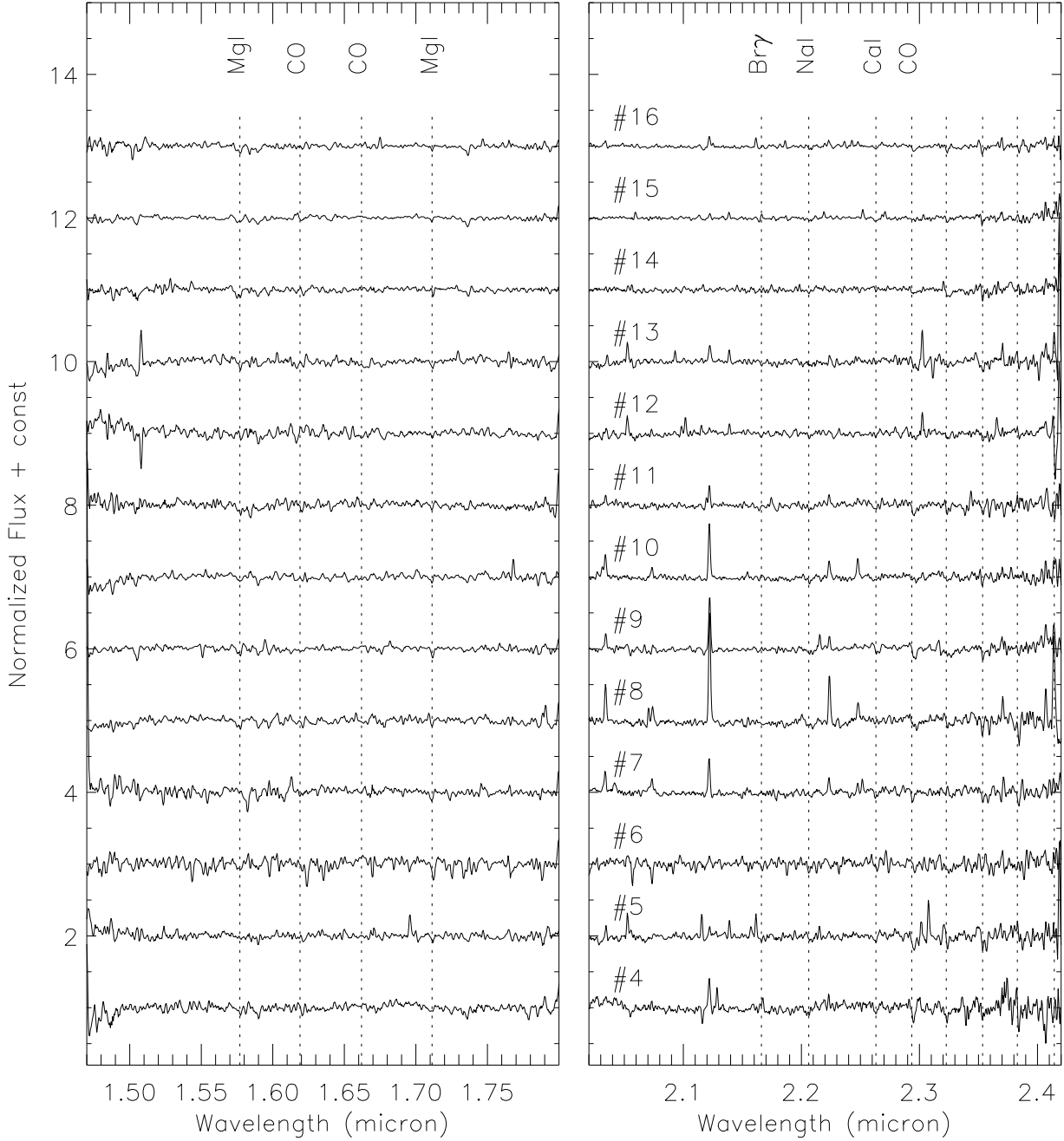}
    \caption{Normalized SINFONI $H$-band (left panel) and $K$-band (right panel) spectra of the 11 objects showing late-type stellar spectral type. And most of them are PMS stars. The vertical lines show the location of some of the absorption lines used for the classification of the stars. The stars are numbered according to Table \ref{spectype}}
\label{ltspec}
\end{figure*}

Besides the 3 B/A stars we found 13 stars showing absorption lines typical of later spectral type (Figure \ref{ltspec}). The most prominent lines we used are the CO first overtone absorption bands between 2.29 and 2.45 $\mu$m, and absorption lines of Mg {\tiny I} and other atomic species in the $H$-band as well as Ca {\tiny I} and Na {\tiny I} in the $K$-band. To classify the late-type stars we use the reference spectra of \citet{cushing2005} and \citet{rayner2009}. The atomic lines, such as Mg {\tiny I} and Na {\tiny I}, are used for the temperature determination, while the CO lines are used to determine the luminosity class. See \citet{bik2010} for a detailed description. 

Compared with the reference spectra,  the CO (2.29 $\mu$m) absorption of our SINFONI spectra is usually deeper than  in dwarf reference spectra, but not as deep as in the giant spectra. In a few cases where a luminosity class IV reference spectrum was available, the spectrum provided a better match to the observed spectrum. This suggests that our late type stars are low- and intermediate PMS stars, and indeed, \citet{luhman1999} and \citet{winston2009} find that PMS spectra have a surface gravity intermediate between giant and dwarf spectra. If the stars were giants, dust veiling could make the K-band CO lines weaker. However, the H-band CO and OH lines are also much weaker, while the atomic lines have the expected EWs. Therefore, this seems to be a surface gravity effect, suggesting a PMS nature of our late type stars.

The "double blind" procedure was applied during the classification of the late type stars to check the accuracy of the classification. In this procedure, Y. Wang and A. Bik did their own classification separately and independently without knowing results of each other, and their results were compared afterwards to get the differences which are the errors of the spectral type in Table \ref{spectype}. Our classification results showed an error of 1 to 2 subclass in spectral type.

The relation from \citet{kenyon1995} was used to convert the spectral type into effective temperature. However, this relation applies only for main sequence stars, for PMS stars a different relation may hold \citep{hillenbrand1997, winston2009}. \citet{cohen1979} show that the temperature of PMS stars might be overestimated by values between 500 K (G stars) and 200 K (mid-K). We took this source of error into account when calculating the errors in the effective temperature.

With the knowledge of the spectral type, we can estimate the extinction from the observed color. As the intrinsic $J-K$ color for dwarfs and giants can differ as much as 0.4 for late K stars \citep{koornneef1983}, for the PMS stars, we used as the intrinsic $J-K$ color the mean of the color for dwarfs and giants. The difference between the mean value and the dwarf and giant values is used as the error in the intrinsic color. The derived extinctions are listed in Table \ref{spectype}.

\begin{table*}
\caption{Photometric and spectroscopic properties of the detected stars brighter than $K=14$ mag.}
\centering
\renewcommand{\footnoterule}{}  % to avoid a line before footnotes
\begin{tabular}{lcccccccc}
\hline \hline
\noalign{\smallskip}
Star  &R.A. (J2000) & Dec. (J2000)  &$K$&$J-K$& T$_{\rm eff}$  &  Sp. Type  &Lum.class &A$_V$\\
~&(h m s)&($^{\circ}$ $^{\prime}$ $^{\prime\prime}$)&(mag)&(mag)&(K)&~&~&(mag)\\
\noalign{\smallskip}
\hline
\noalign{\smallskip}
        1&06:12:54.91&17:59:21.05&11.26$\pm$0.03&1.7$\pm$0.06&15620$\pm$2850&KB4-B7(B3-B7)&V&10.0$\pm$0.3\\
       2&06:12:55.06&17:59:28.93&10.60$\pm$0.02&0.8$\pm$0.03&9800$\pm$1590&KB8-A3(B8-A3)&V& 4.8$\pm$0.2\\
       3&06:12:54.68&17:59:32.82&11.68$\pm$0.04&2.2$\pm$0.09&9800$\pm$1590&KB8-A3(B8-A3)&V&12.8$\pm$0.4\\
       4&06:12:54.82&17:59:12.98&13.39$\pm$0.08&2.3$\pm$0.21&3650$\pm$271&M1.5$\pm$1&V& 8.2$\pm$1.0\\
       5&06:12:53.60&17:59:28.18&13.40$\pm$0.08&2.9$\pm$0.26&4590$\pm$460&K4$\pm$1&V&12.3$\pm$1.2\\
       6&06:12:55.31&17:59:15.90&13.17$\pm$0.07&1.7$\pm$0.20&4730$\pm$446&K3$\pm$1&PMS& 5.4$\pm$1.1\\
       7&06:12:54.59&17:59:17.23&12.57$\pm$0.05&2.2$\pm$0.19&4730$\pm$446&K3$\pm$1&PMS& 8.6$\pm$1.0\\
       8&06:12:54.13&17:59:29.18&13.37$\pm$0.08&2.6$\pm$0.23&4900$\pm$492&K2$\pm$1&V&11.1$\pm$1.1\\
       9&06:12:56.38&17:59:32.75&13.05$\pm$0.07&4.0$\pm$0.39&4900$\pm$492&K2$\pm$1&PMS&19.1$\pm$1.6\\
      10&06:12:54.55&17:59:02.76&12.78$\pm$0.06&3.8$\pm$0.33&5080$\pm$513&K1$\pm$1&PMS&18.2$\pm$1.4\\
      11&06:12:55.10&17:59:21.34&13.14$\pm$0.07&2.3$\pm$0.24&5080$\pm$513&K1$\pm$1&PMS&10.0$\pm$1.2\\
      12&06:12:53.17&17:59:05.93&12.50$\pm$0.05&3.0$\pm$0.25&5385$\pm$630&G8/K0$\pm$2&PMS&14.4$\pm$1.3\\
      13&06:12:55.06&17:59:31.09&12.24$\pm$0.05&1.5$\pm$0.20&5630$\pm$470&G7$\pm$1&PMS& 6.0$\pm$1.2\\
      14&06:12:54.90&17:59:40.96&12.33$\pm$0.05&4.7$\pm$0.42&5385$\pm$630&G8/K0$\pm$2&PMS&24.0$\pm$1.7\\
      15&06:12:54.45&17:59:37.10&13.97$\pm$0.11&2.4$\pm$0.29&4730$\pm$446&K3$\pm$1&Giant& 9.1$\pm$1.4\\
      16&06:12:53.29&17:59:21.70&13.37$\pm$0.08&3.0$\pm$0.27&5250$\pm$519&K0$\pm$1&S. Giant&13.9$\pm$1.2\\
      17&06:12:54.18&17:59:47.58&12.65$\pm$0.00&4.9$\pm$0.00&...&...&Br$\gamma$&...\\
      18&06:12:54.15&17:59:34.19&12.29$\pm$0.00&5.3$\pm$0.00&...&...&Br$\gamma$&...\\
      NIRS 1&06:12:53.83&17:59:23.71&11.36$\pm$0.00&6.2$\pm$0.00&...&...&...&...\\
      NIRS 3&06:12:54.01&17:59:23.68&12.56$\pm$0.00&5.0$\pm$0.00&...&...&...&...\\
\hline
\end{tabular}
\label{spectype}
\end{table*}

\subsubsection{YSOs}
\label{ysos}
\begin{figure*}[htbp]
   \centering
    \includegraphics[angle=0,width= \textwidth]{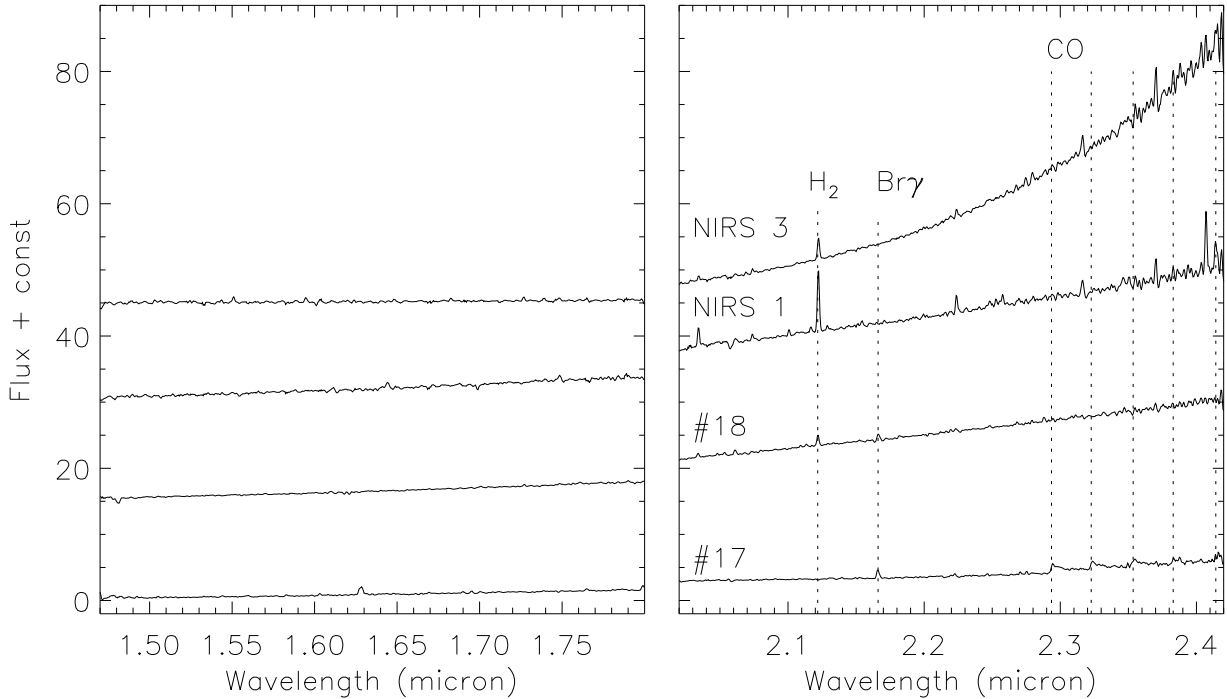}
    \caption{SINFONI $H$-band (left panel) and $K$-band (right panel) spectra of the 4 YSOs. The stars are numbered according to Table \ref{spectype}.}
\label{ysospec}
\end{figure*}

Source \#17 and \#18 have very red spectra and show strong Br$\gamma$ point emission, but the spectra are featureless due to the strong dust emission, which indicates that they are very young and with ongoing accretion activity (Figure \ref{ysospec}). Source \#17 also shows strong $K$-band CO line emissions \citep{bik2004}, which indicates the existence of circumstellar disk. The two infrared sources NIRS3 and NIRS1 which coincide with our mm sources S255IR-SMA1 and S255IR-SMA3, respectively, also have extremely red spectra, only several H$_2$ lines can be seen (Figure \ref{ysospec}). Therefor we cannot get the spectral type of these sources.

\subsubsection{Cluster membership and the HR diagram}
\label{member_hrd}
For the late type stars, as discussed in Sec. \ref{classification}, most of them have a surface gravity typical for PMS stars, so they are considered to be the cluster members. Source \#15 and \#16 show a luminosity type Giant and Super Giant, they are unlikely to be associated with this young cluster and since they show quite high extinction, they are considered to be background stars. 

\begin{figure}[htbp]
   \centering
    \includegraphics[angle=0,width= 8cm]{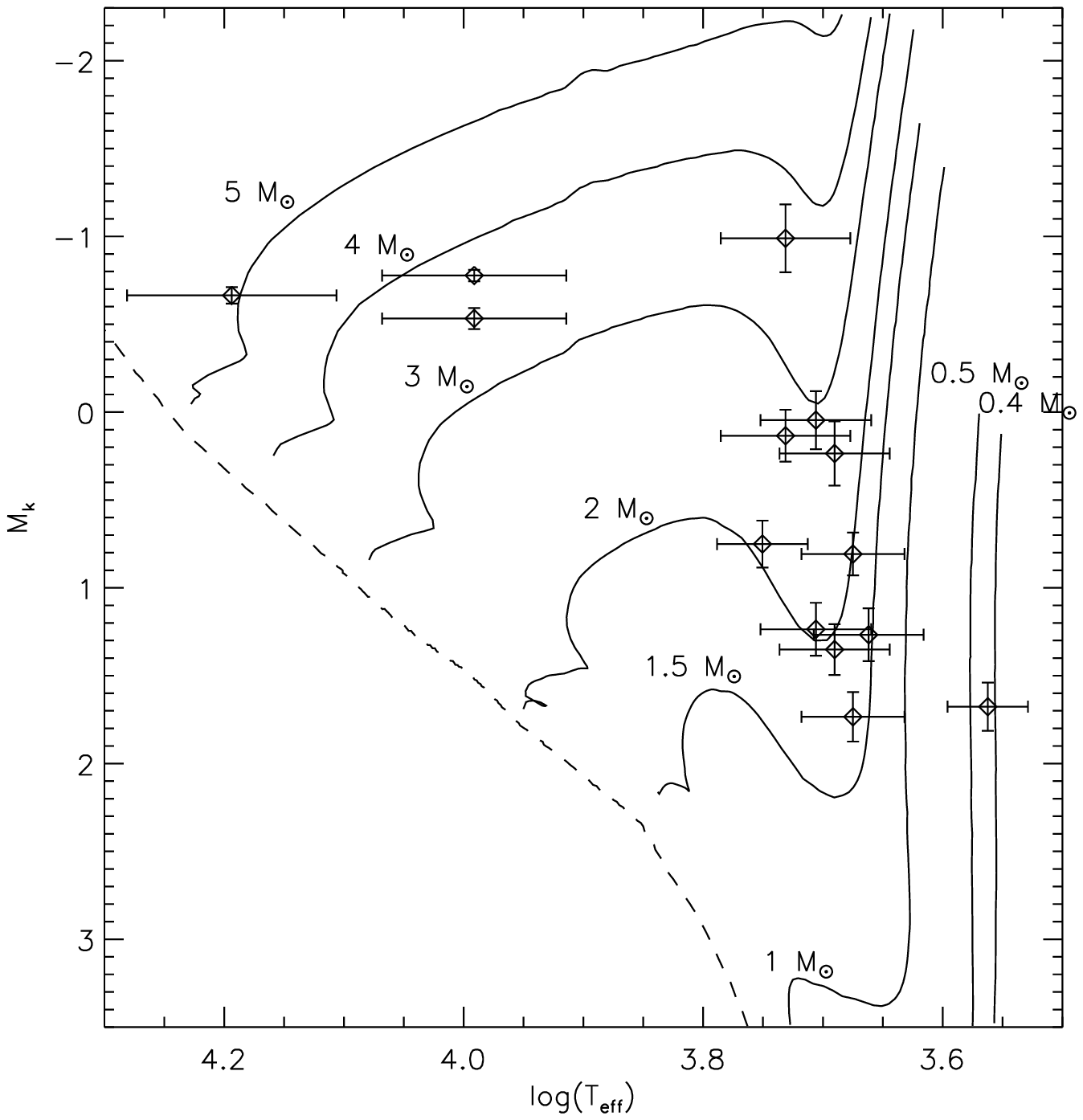}
    \includegraphics[angle=0,width= 8cm]{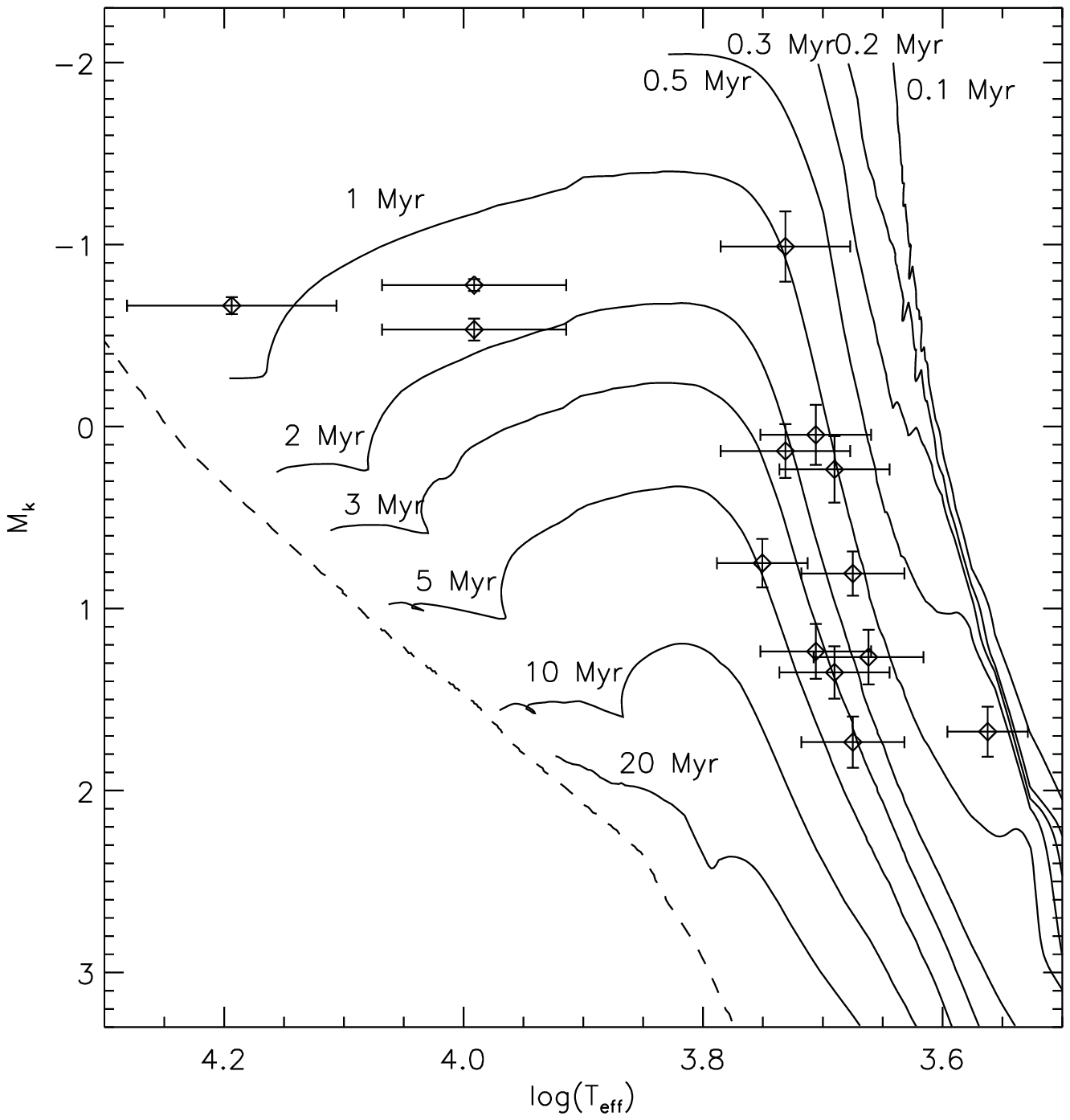}
    \caption{$Top:$ The extinction corrected $K$ vs. log(T$_{\rm eff}$) HRD. The $K-$band magnitude has been corrected for the distance modulus. Over-plotted with a dash line is the 2 Myr main sequence isochrone from \citet{lejeune2001}, and with the solid lines the pre-main-sequence evolutionary tracks from \cite{dario2009}. $Bottom:$ The same data but over-plotted with solid lines showing the isochrones from \cite{dario2009}.}
\label{hrd}
\end{figure}

To constrain the mass of the cluster members and the age of the whole cluster, we construct a HRD to compare the observed parameters with PMS evolutionary tracks. The derived extinctions allow us to de-redden the $K$-band magnitude, convert to absolute magnitude with the derived temperature, and plot the points in a $K$ vs. log(T$_{\rm eff}$) diagram (Fig. \ref{hrd}, top). We exclude the background stars \#15 and \#16 to enable a comparison with the isochrones. The over-plotted dashed line is the 2 Myr main sequence isochrone taken from \citet{lejeune2001}. The solid lines in the left panel represent the evolutionary tracks taken from \citet{dario2009} using the models of \citet{siess2000} for 8 different masses: 5, 4, 3, 2, 1.5, 1, 0.5, 0.4 M$_{\odot}$. Comparison of the location of the PMS stars with these evolutionary tracks yields an approximate mass varying from $\sim$ 5 M$_{\odot}$ for the brightest stars to 0.4 M$_{\odot}$.

In the bottom panel of Figure \ref{hrd}, the over-plotted solid lines are the PMS isochrones between 0.1 and 10 Myr. The location of the stars is consistent with a range of age. Besides source \#1, \#14 and \#4 which lie between the 0.5 and 1 Myr isochrones, most of the objects span the $1-3$ Myr isochrones, with the more massive objects closer to the 1 Myr isochrone. For the less massive objects, the spread in age is larger, most likely due to higher uncertainties in the spectral type determination. The location of the PMS stars suggests an age of $2\pm1$ Myr.

Comparison of the location of the PMS stars in the HRD with those of the Herbig AeBe stars \citep{vandenancker1998} shows that the late spectral type PMS stars are younger than the Herbig AeBe stars and will evolve from their present G- and K spectral type to late B, A or early F spectral type when they become main sequence stars. The three relative early spectral type sources (\#1, \#2 and \#3) already evolved to late B early A spectral type. They are in the transit phase between our late spectral type sources and the main sequence. 

\begin{figure}[htbp]
   \centering
    \includegraphics[angle=-90,width= 8cm]{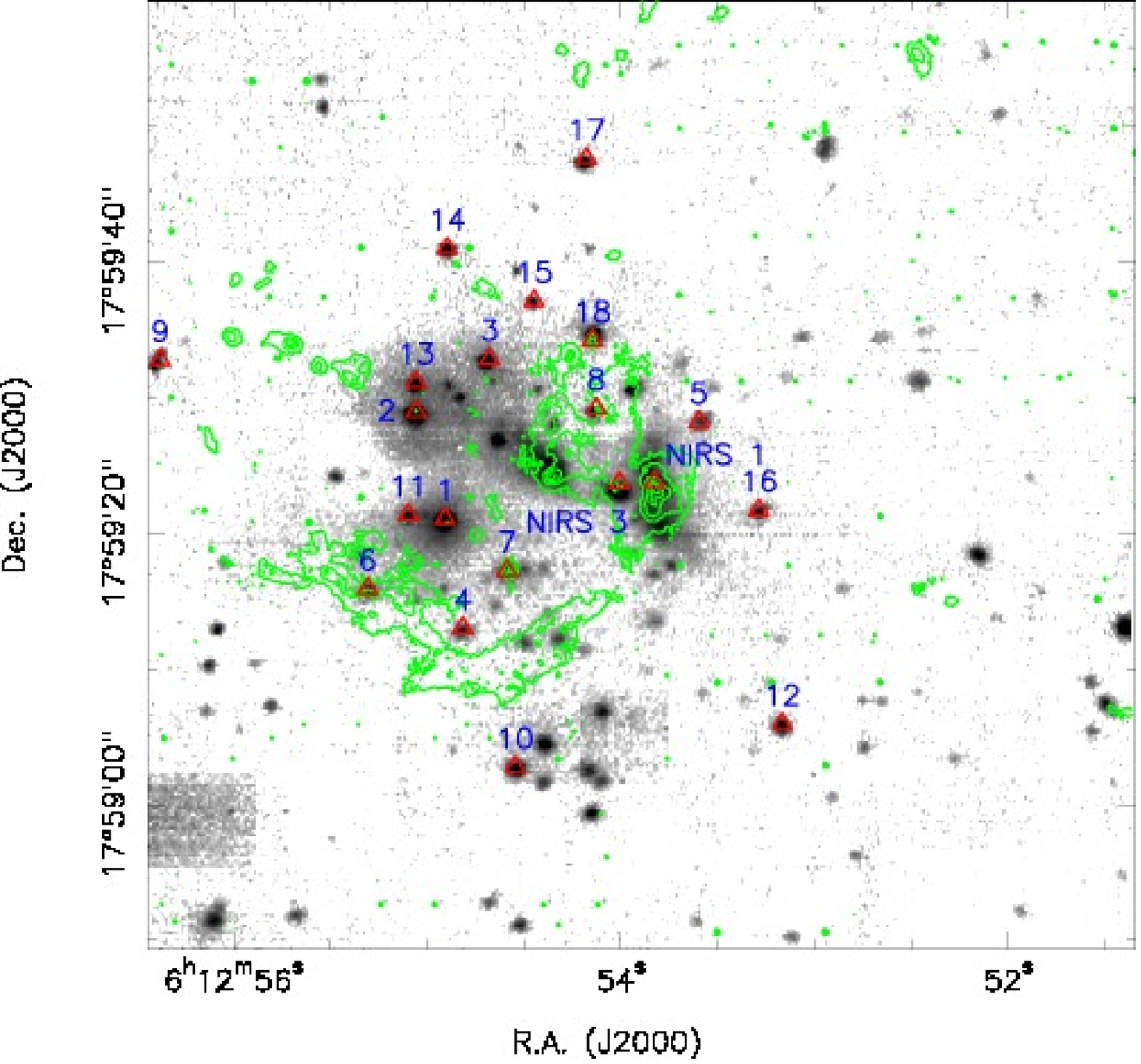}
    \caption{The stars listed in Table \ref{spectype} marked on Br$\gamma$ line$+$continuum map overlaid with H$_2$ line emission contours.}
\label{position}
\end{figure}

\section{Discussion}

\subsection{Different evolutionary stages and triggered star formation?}
\label{diff_pop}
The SMA and IRAM 30m data together reveal three massive star formation regions with different evolutionary stages. The SCUBA 850 $\mu$m image (Figure \ref{spitzer}) presents three continuum peaks in the whole S255 complex region, one in each sub-region, which is S255N, S255IR, S255S, from north to south respectively. With our high resolution SMA observations, $\sim$2500 AU at the given distance of 1.59 kpc, we found that all SCUBA 850 $\mu$m sources fragment into several cores. 

\citet{minier2007} suggests S255S to be at a very young stage without active star formation, however, our observations show outflows associated with the mm sources, Furthermore, the virial mass is smaller than the mass obtained from the SCUBA 850 $\mu$m measurement, which implies that the S255S region will likely collapse. Since the peak column density is also on the order of the proposed threshold for high-mass star formation of 1 g cm$^{-2}$ \citep{krumholz2008}, S255S is at a very early stage of ongoing star formation and may form massive stars. The single dish continuum properties of S255IR and S255N do not show much difference, however, our interferometry and NIR observations show us the different properties of S255IR and S255N. While the large scale NIR emission shows the existence of the NIR cluster in S255IR and not so many NIR point sources in S255N, which indicate S255IR is most likely in a more evolved stage compared to S255N. For the individual sub-cores, more lines are detected at S255IR-SMA1 than S255N-SMA1, which indicates a higher temperature and more evolved nature of S255IR-SMA1 compared to S255N-SMA1. Regarding the kinematic properties, S255N-SMA1 has similar outflow velocities but a much smaller size of the outflow than S255IR-SMA1, which may also suggest a younger evolutionary stage of S255N-SMA1. Among other SMA mm continuum cores, S255IR-SMA2, S255N-SMA2 and S255N-SMA3, which do not have many line emissions are considered to be at earlier evolutionary stages.

Figure \ref{spitzer} presents the whole star formation region. The young star formation region S255 complex lies between the evolved H {\tiny II} regions, S255 and S257. From the morphology of the dust structure and the H {\tiny II} regions,  it appears that the two H {\tiny II} regions pushed gas between them and formed the S255IR dust structure and triggered the star formation, which has also been suggested by \citet{bieging2009,minier2007}. To inquire that, we studied the velocity map of our 30m data, however, due to the high noise level at the edge of the map, we could not find a significant velocity difference between the west and east edge of the dust structure to prove the interaction between the H {\tiny II} regions and the dust structure. \citet{chavarria2008} estimated the dynamical age of the two H {\tiny II} regions S255 and S257, which  is $\sim1.5\times10^6$ yr, similar to our NIR cluster in S255IR. So we can not prove the triggering star formation. However, we witness outflow interaction between S255IR and S255N (Sec. \ref{outflow}). Since S255N is younger compared to S255IR, S255N may be affected by S255IR.

Further more, the age of the cluster around S255IR we obtained from the SINFONI data is 2$\pm$1 Myr (Sec. \ref{member_hrd}). \citet{chavarria2008} obtained an age of the larger-scale cluster of 1 Myr which is consistent with our result. However, the dynamical age of the high-mass protostar in this region, S255IR-SMA1, is $\sim10^4$ yr (Table \ref{outflow_p}). Without the knowledge of the outflow inclination angle, we could systematically underestimate the age value by a factor of 2 to 5 \citep{cabrit1992}. Therefore, S255IR-SMA1 should not be dynamically older than 10$^5$ yr and S255IR-SMA2 is even younger than S255IR-SMA1. The age difference between the NIR cluster and the massive protostellar objects indicates that the most massive sources in the cluster form last.

\citet{minier2007} suggests a collect-and-collapse and triggered star formation scenario for S255IR, which is that the B-stars (from our observations they are late B stars, i.e. star \#1, \#2, \#3 in Figure \ref{position}) in S255IR formed through a collect-and-collpase process, and triggered the formation of NIRS 1 and NIRS 3. The NIR and mm data presented in this paper show that the cluster detected in the infrared is likely older than the mm sources detected with our SMA data. While the SMA mm sources lie in the center of the NIR cluster, the NIR cluster members are more distributed around the mm sources. Based on this information, we propose an outside-in star formation scenario, which is that the central gas filament collapses under the pressure of the two H {\tiny II} regions. The collapse of the clump may start outside-in under the pressure of the two H {\tiny II} regions, the low mass cores need lower density to form, and they formed first at the outside region of the clump and collapse to stars. And then either the low- to intermediate- mass stars may enhance the instability of the central high-mass cores and  potentially trigger the high-mass star formation in S255IR, or the massive cores could build up slightly slower and start to collapse afterwards forming the most massive stars in this region. Our outflow observations show a hint that the energetic outflow from the YSOs in S255IR may again have affected the star formation in S255N, but this needs further observations to confirm.

However, the different evolutionary stages of the various regions and even within each region are quite clear. This suggests a sequential star formation.

\subsection{Multiple outflows}
We detected outflows in all three regions (Sec. \ref{outflow}). In the top panels of Figure \ref{s255iroutcom}, the outflow emission which is nearby the SMA continuum sources does not really follow the NE-SW direction. We suggest this complicated outflow environment in the nearby region of the continuum sources is due to the interaction between outflows from NIRS 1 and NIRS 3. The red-shifted lobe to the south of the continuum source (bottom panels in Figure \ref{s255iroutcom}) is most likely driven by NIRS 1, because the north-south bipolar reflection nebular which is associated with NIRS 1 follows this direction to the south \citep{jiang2008, simpson2009}, and the outflow driven by NIRS 3 should be blue-shifted at this direction. 

In Figure \ref{s255noutcom}, the red-shifted gas at the bottom part of line (b) seems to be associated with S255N-SMA3, however, another blue-shifted feature at the southern part of the map shows up as the resolution changes. Figure \ref{s255_out} shows that these components are mixed together with the outflows in S255IR, which may indicate the interaction between these two regions. 

In the youngest region S255S, the velocity of the outflows is much smaller compared to the other two regions. Table \ref{scubacont} shows that the column density of this region is still relatively low, S255S is in very young evolutionary stage and likely just start collapsing. This is the likely reason we did not detect very energetic outflows like for the other two regions in this region.

\subsection{Disk candidates in S255IR}
\label{disk_can}
NIRS 1, which coincides with S255IR-SMA3, is reported to have a polarization disk \citep{jiang2008}, \citet{simpson2009} also reported a disk, however, with a slightly different interpretation. Because NIRS 1 is relatively more evolved, we detected only one unresolved mm continuum source associated with only the CO isotopologue lines (Sec. \ref{Spectral}). NIRS 3, which coincides with S255IR-SMA1, is considered to be a high-mass protostar based on several signatures (e.g., UCH{\tiny II} region, maser emissions, hot core emissions and energetic outflows, see Sec. \ref{introduction} and \ref{results}). We detected a rotational structure coinciding with this source perpendicular to the outflow (Figure \ref{s255ir_mom1}). The HCOOCH$_3$ position velocity diagram shows that this rotational structure is not Keplerian, so it could be a rotating toroid around NIRS 3. However, the C$^{18}$O pv-diagram shows a much larger Keplerian-like velocity structure perpendicular to the outflow (Figure \ref{s255ir_mom1_pv}). If this source were at a much further distance or observed with worse resolution, this structure could easily be identified as a disk. However, with our resolution we resolved two mm source, and we know it is not a disk. The C$^{18}$O gas size is $\sim2\times10^4$ AU and the continuum source has a size of 1.6$\times10^4$ AU at the given distance of 1.59$\times10^3$ pc, which is similar to the rotational structure described in \citet{fallscheer2009}. This structure is also much larger than the jeans length which is $\sim$6 000 AU for a temperature  $\sim$20 K and a density $\sim1\times10^6$. Our system also exceeds the criteria to maintain a stable disk described by \citet{kratter2010}. Therefore, we suggest this structure to be a rotating toroid which fragmented into several sources, and form a multiple system. Similar structures of fragmenting pseudo-disks have also been modeled by \citet{krumholz2009} and \citet{peters2010}. 

Following these, we propose a star formation scenario for this large rotational structure, which is that the rotational elongated dust structure formed first, and at the central region the massive source NIRS 1 started forming. However, this large structure is much more massive compared to NIRS 1 and is unstable, then it fragmented into two sources, S255IR-SMA1 and S255IR-SMA2.

The SINFONI source \#17 shows strong Br$\gamma$ and CO emission in $K$-band (Figure \ref{ysospec}), which indicates the existence of a circumstellar disk. Source \#17 might also drive the jet like emissions at the northern edge of Figure \ref{three-color}. It is interesting to find disk signatures toward sources with that different evolutionary stages within the same forming cluster.

\subsection{Cores and clumps}
Since the single dish has a much larger beam, which can cover all the continuum sources, and $^{13}$CO may be optically thick, the observed velocity traces the outer layer of the whole clump and shows the mean line-of-sight velocity. The molecular lines we used to get the velocities for the SMA observations are all dense gas and disk tracers \citep{cesaroni2007}, so what we obtained are the velocities of the high mass cores (Table \ref{vlsr}). The difference between the core velocities and the clump velocities of several km s$^{-1}$ (Table \ref{vlsr}) are very different from the low-mass core cases (e.g. 0.46 km s$^{-1}$ in NGC 1336 \citep{walsh2007}, 0.17 km s$^{-1}$ in Oph A \citep{difrancesco2004}). This difference confirms that massive star formation regions are more turbulent than low-mass ones. It further implies stronger peculiar motions of the protostars and cores within the clump/cluster gravitational potential.

\section{Summary}
Combining multi-wavelength data from mm to NIR wavelength, we characterize the different (proto) stellar populations within the S255 star formation complex. S255S, S255N and S255IR show different dynamical and chemical properties, not only at mm wavelength but also at infrared wavelengths, which indicates they are in different evolutionary stages. With the SMA, IRAM 30m and VLT-SINFONI observations, we found outflows in all three regions, high velocity collimated ones in S255IR, high velocity more confined ones in S255N and lower velocity confined ones in S255S. The multiple outflows we found in S255IR and S255N suggests a potential interaction between these two regions. From the outflow maps, we estimated the dynamical age of the outflow driving sources. Although without the information of the inclination angle this dynamical age could be underestimated by a factor of 2 to 5 \citep{cabrit1992}, our mm sources should not be older than 10$^5$ yr.

We detected 25 molecular lines in S255IR, including 7 lines of the CH$_3$CN(12$_k-11_k$) k-ladder with k = 0...6, confirming that the hot core nature of S255IR-SMA1. Only 10 molecular lines are detected in S255N, including 2 CH$_3$CN(12$_k-11_k$) k-ladder with k = 0, 1, which is indicative of a younger age and colder temperature. Besides the 3 CO isotopologue line emissions, only diffuse SO emission is detected in S255S. This is consistent with different chemical ages.

High-density tracers like CH$_3$CN and HCOOCH$_3$ show rotational structures around the most prominent high-mass protostar candidates in S255IR and S255N. Furthermore in S255IR, the C$^{18}$O presents an elongated rotational structure with a Keplerian-like velocity gradient perpendicular to the outflow. With a size of $\sim2\times10^4$ AU, this structure can not be a disk but may be a rotational toroid which fragments into several sources.

Near infrared $H$- and $K$-band integral field spectroscopy observations were done for S255IR. We identified the excitation mechanism of the H$_2$ emission. We derived the spectral type of 16 stars, and 14 of them are considered to be the cluster members. With the knowledge of the spectral type and the SOFI $J$- and $K$- band photometry results \citep{bikthesis} of the cluster members, we constructed an HR diagram to estimate the age of the cluster, which is 2$\pm$1 Myr . This age is consistent with the result \citet{chavarria2008} obtained. The age difference between the low-mass cluster and the massive mm cores indicates different stellar populations in the cluster. This also leads to a question, do the massive stars in this cluster form last? Our data support the idea that massive stars form last. We propose the triggered outside-in collapse star formation scenario for the bigger picture and the fragmentation scenario for S255IR

The whole picture of the S255 complex suggests triggered star formation, however, but we can not give hard proof at present. However, the different evolutionary stages between each region and different stellar populations in S255IR are consistent with sequential star formation.

\begin{acknowledgements}
The authors thank C. J. Cyganowski for providing the VLA 3.6 cm data of S255N, F. Eisenhauer for providing the data reduction software SPRED, A. Modigliani for help in the data reduction, R. Davies for providing his routines to remove the OH line residuals and N. Da Rio for providing the isochrones and evolutionary tracks, and M. Fang for the discussion.
 
 Y.W. acknowledges support by Purple Mountain Observatory, CAS and Max-Plank-Institute for Astronomy. E. P. is funded by the Spanish MICINN under the Consolider-Ingenio 2010 Program grant CSD2006-00070: First Science with the GTC.
 \end{acknowledgements}

\bibliographystyle{aa}
\bibliography{wang2010}{}

\end{document}